\documentclass[twocolumn]{aastex63}

\usepackage{newtxtext,newtxmath}
\usepackage{threeparttable}
\usepackage[T1]{fontenc}
\usepackage{ae,aecompl}
\usepackage[english]{babel}
\usepackage{natbib}
\usepackage{amsmath,amsfonts,amssymb,textcomp}
\usepackage{amssymb}
\usepackage{varioref}
\usepackage{hyperref}
\usepackage{psfig}
\usepackage{float}
\usepackage{graphicx,graphics}
\usepackage[caption=false]{subfig}
\hypersetup{colorlinks=true,citecolor=blue, urlcolor=magenta}

\received{}
\revised{}
\accepted{}
\submitjournal{ApJ}

\shorttitle{LoCuSS: SPLASHBACK RADIUS DETECTION}
\shortauthors{Bianconi et al.}

\begin{document}
\graphicspath{{images/}}

\title{LoCuSS: The splashback radius of massive galaxy clusters and its dependence on cluster merger history}

\correspondingauthor{Matteo Bianconi}
\email{mbianconi@star.sr.bham.ac.uk}
\author[0000-0002-0427-5373]{Matteo Bianconi}
\affiliation{School of Physics \& Astronomy, University of Birmingham, Birmingham, B15 2TT, UK
}

\author[0000-0002-7387-6754]{Riccardo Buscicchio}
\affiliation{School of Physics \& Astronomy, University of Birmingham, Birmingham, B15 2TT, UK
}
\affiliation{Institute for Gravitational Wave Astronomy, University of Birmingham, Birmingham, B15 2TT, UK}

\author[0000-0003-4494-8277]{Graham P. Smith}
\affiliation{School of Physics \& Astronomy, University of Birmingham, Birmingham, B15 2TT, UK
}

\author[0000-0003-3255-3139]{Sean L. McGee}
\affiliation{School of Physics \& Astronomy, University of Birmingham, Birmingham, B15 2TT, UK
}

\author[0000-0003-3255-3139]{Chris P. Haines}
\affiliation{Instituto de Astronom\'ia y Ciencias Planetarias de Atacama, Universidad de Atacama, Copayapu 485, Copiap\'o, Chile}

\author[0000-0002-4606-5403]{Alexis Finoguenov}
\affiliation{Department of Physics, University of Helsinki, Gustaf H\"allstr\"omin katu 2a, FI-0014 Helsinki, Finland}

\author[0000-0003-1746-9529]{Arif Babul}
\affiliation{Department of Physics and Astronomy, University of Victoria, Victoria, BC V8P 1A1, Canada}

\begin{abstract}
We present the direct detection of the splashback feature using the sample of massive galaxy clusters from the Local Cluster Substructure Survey (LoCuSS). This feature is clearly detected (above $5\sigma$) in the stacked luminosity density profile obtained using the K-band magnitudes of spectroscopically confirmed cluster members. 
We obtained the best-fit model by means of Bayesian inference, which ranked models including the splashback feature as more descriptive of the data with respect to models that do not allow for this transition.  In addition, we have assessed the impact of the cluster dynamical state on the occurrence of the splashback feature. We exploited the extensive multi-wavelength LoCuSS dataset to test a wide range of proxies for the cluster formation history, finding the most significant dependence of the splashback feature location and scale according to the presence or absence of X-ray emitting galaxy groups in the cluster infall regions. In particular, we report for the first time that clusters that do not show massive infalling groups present the splashback feature at a smaller clustercentric radius $ r_{\rm{sp}}/r_{\rm{200,m}} = 1.158 \pm 0.071$ than clusters that are actively accreting groups  $r_{\rm{sp}}/r_{\rm{200,m}} =  1.291 \pm 0.062$.  The difference between these two sub-samples is significant at $4.2\sigma$, suggesting a correlation between the properties of the cluster potential and its accretion rate and merger history. Similarly, clusters that are classified as old and dynamically inactive present stronger signatures of the splashback feature, with respect to younger, more active clusters. We are directly observing how fundamental dynamical properties of clusters reverberate across vastly different physical scales.

\end{abstract}


\section{Introduction} \label{sec:intro}

\begin{figure*}
\centering
\includegraphics[width=0.39\linewidth, keepaspectratio]{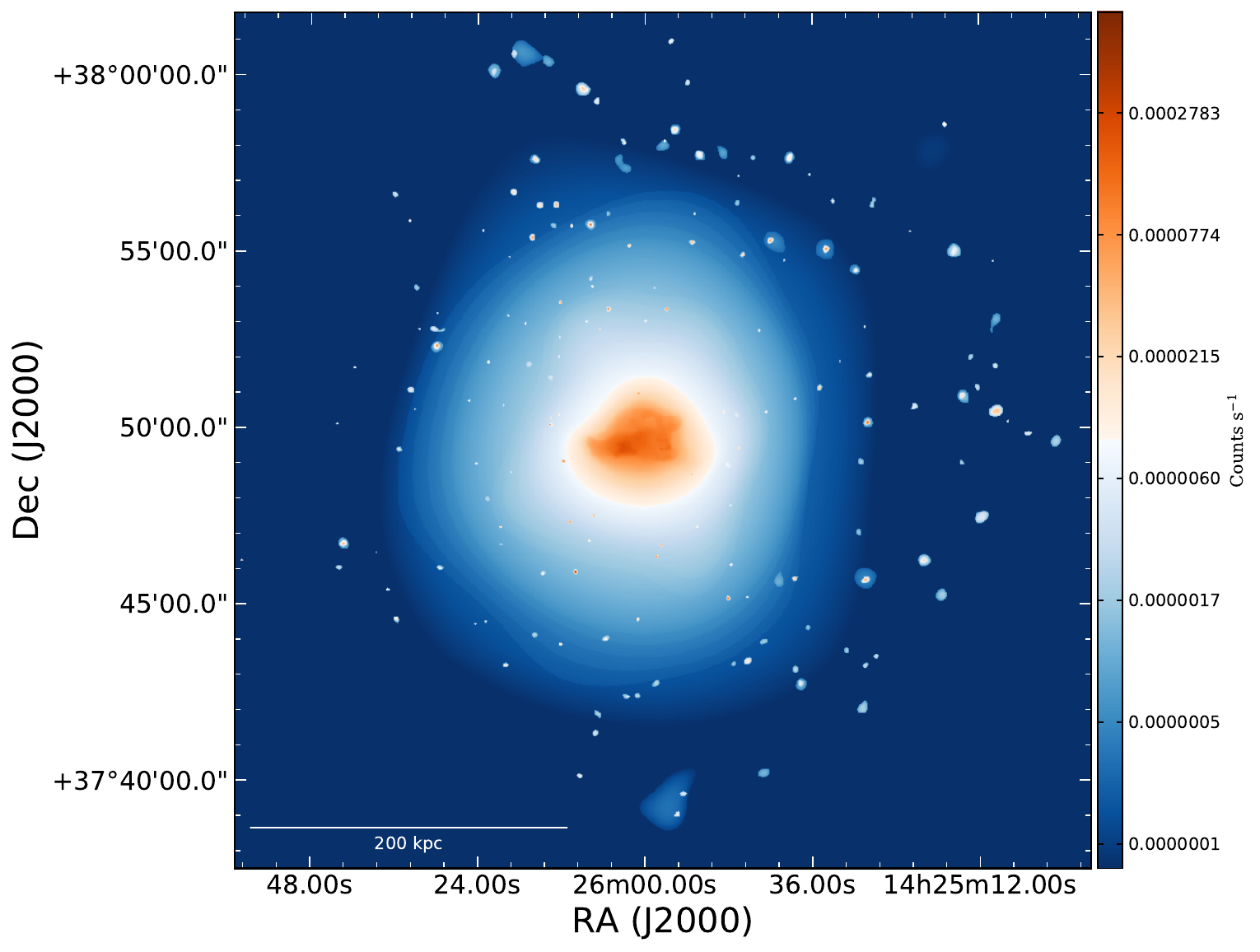}\includegraphics[width=0.3\linewidth, keepaspectratio]{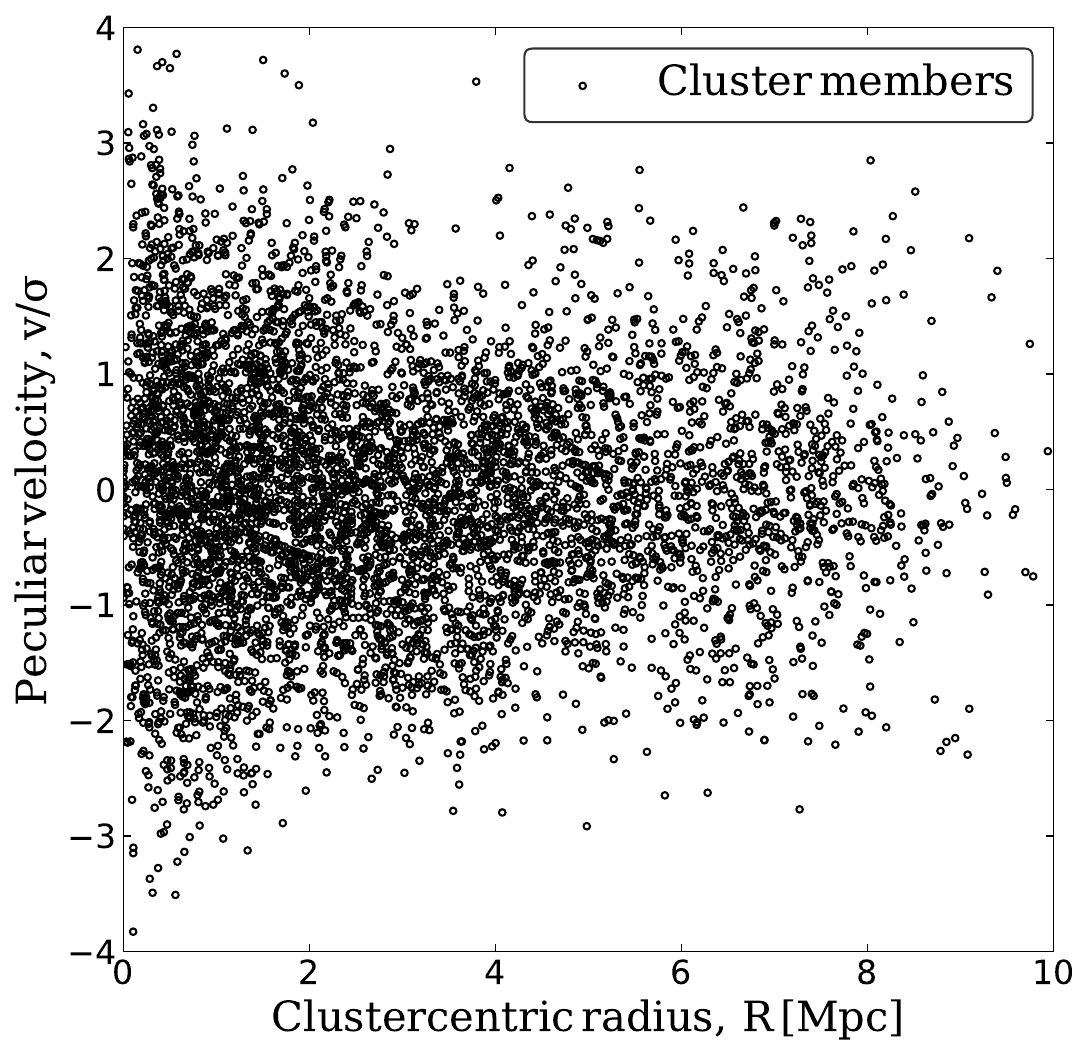}\includegraphics[width=0.3\linewidth, keepaspectratio]{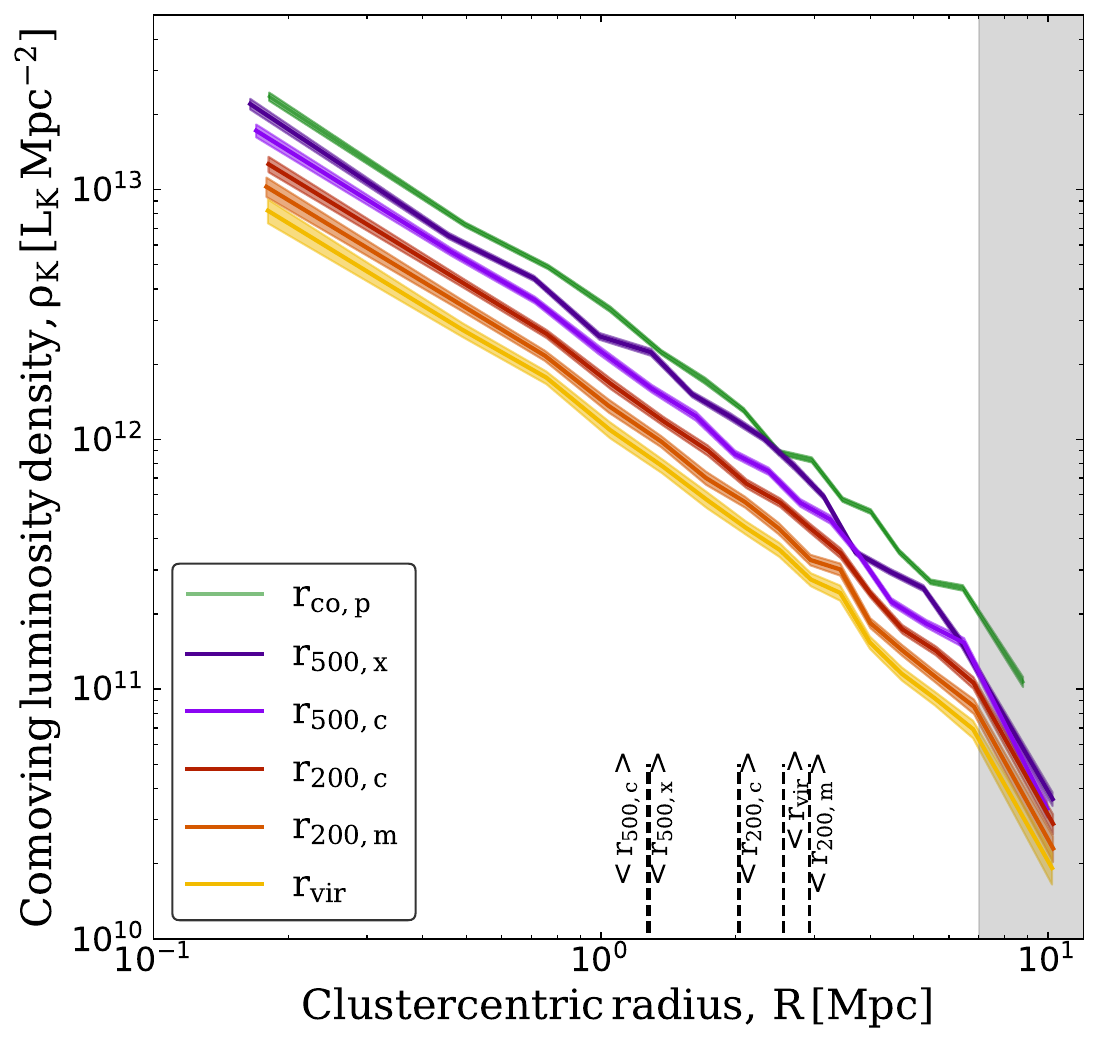}
\caption{Left panel: X-ray emission $\rm counts\,s^{-1}$, tracing the diffuse hot intracluster medium which permeates the central region of cluster Abell 1914, and individual emission from active galactic nuclei, observed using the \textit{Chandra} telescope. Middle panel: stacked phase-space diagram of the cluster members selected following the procedure outlined in Section~\ref{sec:dataset}. The clustercentric radius of spectroscopically confirmed cluster members of 20 LoCuSS clusters is plotted with respect to their peculiar velocity relative to the central redshift of the their host cluster, scaled by the velocity dispersion of all the cluster members within $ r_{\rm{200}}.$ The peculiar velocity of each galaxies is computed as following $ v_{\rm{gal}} = c \times (z_{\rm{gal}} - z_{\rm{halo}})/(1 + z_{\rm{halo}})$, where  $z_{\rm{gal}}$ and $z_{\rm{halo}}$ are the galaxy and its parent halo redshift. Right panel: comoving luminosity density of LoCuSS cluster galaxies, plotted with respect to comoving clustercentric  radius. The individual profiles refer to different density threshold scaling, and are artificially offset vertically for viewing purposes except for the profile labelled $r_{\rm{500,c}}$. The profile labelled $ r_{\rm{co,p}}$ marks the luminosity profile without any scaling. The values of $r_{\rm{500,c}}$ and $r_{\rm{500,x}}$ are computed using $\rm \Lambda$CDM cosmology \citep{martino14, okabe16}, and are displayed solely for comparison. The grey shaded area marks the radial completeness threshold defined in Section~\ref{sec:dataset}.}\label{rholk_profiles_scaled}
\end{figure*}
The outskirts of galaxy clusters are the new frontier to improve our understanding of the multi-faceted physical processes impacting baryons and dark matter during structure formation. They mark the transition between pristine field sparseness and evolved collapsed haloes.  Here merger shocks are produced and galaxy evolution is jolted, due to accretion on to clusters \citep{walker19}. This makes cluster outskirts the prime targets to observe the transformation of galaxy properties \citep{haines15}, to assess biases in clusters mass estimates \citep{reiprich13}, and test predictions of cosmological models \citep{pratt19}.

The collapse and growth of dark matter haloes follows the evolution of primordial density perturbations whose gravitational pull locally overcomes the expansion of the Universe. In the simplest scenario involving spherically-symmetric, continuous accretion, virialization during collapse redistributes the potential energy of the primordial overdensities, allowing for a final stable structure characterised by a specific density contrast $ \Delta_{\rm{vir}}$ (\citealt{gunn72}, see also \citealt{voit2005} for a review). It follows that the mass enclosed within the virialized sphere is $ M_{\rm{vir}}= \frac{4}{3}\pi r_{\rm{vir}}^3 \Delta_{\rm{vir}} \bar{\delta}$ \citep{cole96}.  This definition is commonly used to indicate gravitationally bound structures, while allowing for different choices of $ \Delta \in [180,200,500]$ and reference density $\rm\bar{\delta}$, namely the mean matter density or critical density of the Universe. This formalism is particularly advantageous since it permits a bridge between simulations and observations, and it has been used to probe self-similarity for a wide range of halo properties, involving both baryonic and non-baryonic matter, in regimes that are dominated by gravity (e.g. \citealt{navarro1996,bullock01,tinker08,schaller15, springel18,farahi19}).

Modifications have been proposed to the idealized collapse model to allow for more general conditions of structure growth, such as those in $\rm \Lambda CDM$ cosmology \citep{bryan98}. However, it is a challenge to fully describe the structure on the outskirts of these halos, with models generally unable to capture the rich extent of density fields that connect haloes which are gravitationally bound despite being separated by many virial radii \citep{prada06}. Indeed, as the density contrast is defined against mean cosmic densities which have a redshift dependence, an intrinsic pseudo-evolution of halo masses arises from the formalism \citep{babul02,diemand07, diemer13b}. Naively, the spherical collapse model suggests the presence of a sharp transition in the radial density profile of the forming halo, corresponding to the region which physically separates gravitationally bound matter accreted previously from newly-infalling regions \citep{fillmore84, bertschinger85, adhikari14, shi16}. Infalling particles which get trapped in the growing halo potential populate a shell at the apocenter of their first orbit -- the so-called splashback region. For this reason, the splashback radius has been proposed as physical boundary of collapsed haloes and is detected as a steepening of the radial density profile of dark matter halos from large-scale numerical simulations (\citealt{diemer14, more15}, see also \citealt{tomooka20}).

Observationally, the challenge of detecting the splashback radius has been recently undertaken using galaxies and gas measured in large-scale surveys as tracers of the underlying dark matter halo potential in galaxy clusters. Unfortunately, it has become apparent that the measured splashback radius can be significantly affected by the cluster selection method. Optical selection methods based on cluster richness measured within an aperture result in an underestimated splashback location due to projection effects and interloper contamination \citep{zu17, busch17, baxter17}. Additional biases on the measure of the splashback location have been confirmed for methods relying on clusters selected using \texttt{redMaPPer} \citep{rykoff14} by comparison with cluster mass profiles obtained using weak-lensing shear \citep{chang18}.  More recently, selection methods which are unaffected by the same biases, and thus minimize the risk of spurious correlations between the splashback radius and the cluster selection, such as Sunyaev-Zel'dovich (SZ, \citealt{sunyaev72}) \citep{shin18, zuercher19, adhikari20} and X-ray  \citep{umetsu17}, have been attempted. To date, these methods have yielded less precise measurements of the splashback radius due to their smaller samples \citep{more15, baxter17, murata20}, and to the typically noisy measurements from weak-lensing at these radii.

\begin{deluxetable*}{l c c c r c c}
\tablecaption{List of clusters selected for this study, presented together with central coordinates, average redshift, weak-lensing mass $ M_{\rm{200,m}}$ from \citet{okabe16}, and membership of the main sub-samples discussed in the text, and based on \cite{haines17} and \cite{sanderson09}. 
\label{table_clusters}}
%\tablewidth{120pt}
%\tabletypesize{\scriptsize}
\tablehead{
Name & RA(J2000) & Dec(J2000) & Redshift & $ M_{\rm{200,m}}$~~~~~ & Infalling & Low entropy\\
& & & $ \langle z \rangle $ & $ 10^{14} h^{-1} M_{\odot}$ & groups? & core? } 
\startdata
Abell 68 & 00:37:06.84	& +09:09:24.28 & 0.251 & $8.39_{-1.64}^{+2.00}$ & Y & N\\
ZwCl\,0104.4+0048 & 01:06:49.50 & +01:03:22.10 & 0.253 & $2.98_{-1.26}^{+2.21}$ & N & Y\\
Abell 209 & 01:31:53.45 & $-$13:36:47.84 & 0.209 & $17.01_{-2.93}^{+3.70}$ & Y & N\\
Abell 383 &	02:48:03.42 &	$-$03:31:45.05 &	0.189 & $6.90_{-1.64}^{+2.18}$ & N & Y\\
Abell 586 & 07:32:20.22 &	+31:37:55.88 &	0.171 & $8.32_{-2.32}^{+3.54}$ & N & N\\	
Abell 611 & 08:00:56.81 &	+36:03:23.40 &	0.286 & $11.77_{-2.35}^{+2.77}$ & Y & N\\
Abell 697 & 08:42:57.58 &	+36:21:59.54 &	0.282 & $14.22_{-3.73}^{+6.14}$ & Y & N\\
ZwCl\,0857.9+2107 & 09:00:36.86 &	+20:53:39.84 &	0.234 & $3.52_{-1.39}^{+1.97}$ & N & Y\\	
Abell 963 & 10:17:03.65 &	+39:02:49.63 &	0.204 & $9.46_{-1.79}^{+2.20}$ & Y & Y\\
Abell 1689 & 	13:11:29.45 &	$-$01:20:28.32 &	0.185 & $13.15_{-1.97}^{+2.32}$ & N & N\\
Abell 1758 & 	13:32:33.50 &	+50:30:31.61 &	0.279 & $7.22_{-1.83}^{+2.42}$ & Y & N\\
Abell 1763 & 13:35:18.07 &	+40:59:57.16 &	0.232 & $22.89_{-4.32}^{+5.94}$ & Y & N\\
Abell 1835 & 14:00:52.50 &	+02:52:42.64 &	0.252 & $12.27_{-2.28}^{+2.75}$ & Y & Y\\
Abell 1914 & 14:25:59.70 &	+37:49:41.63 &	0.167 & $12.51_{-2.65}^{+3.55}$	& Y & N\\
ZwCl\,1454.8+2233 &	14:57:15.11 &	+22:20:34.26 &	0.257 & $6.28_{-2.69}^{+6.10}$ & Y & Y\\
Abell 2219 & 16:40:22.56 &	+46:42:21.60 &	0.226  & $15.17_{-3.16}^{+4.53}$ & Y & N\\
RXJ\,1720.1+2638  & 17:20:10.14 &	+26:37:30.90 &	0.160 & $7.23_{-2.26}^{+3.46}$ & Y & Y\\
RXJ\,2129.6+0005 & 21:29:39.88 &	+00:05:20.54 &	0.234 & $7.35_{-2.48}^{+4.11}$ 	& N & Y\\
Abell 2390 & 21:53:36.85 &	+17:41:43.66 &	0.229 & $13.75_{-2.42}^{+2.91}$ & Y & Y\\
Abell 2485 &	22:48:31.13 &	$-$16:06:25.60 &	0.247 & $7.56_{-1.74}^{+2.27}$ & N & N\\
\enddata
\end{deluxetable*}
Simulations indicate that cluster outskirts are continuously bombarded by smaller haloes, galaxy- or group-like, which is the prevalent ingredient for the fast evolution of cluster mass, especially below redshift $ z<0.5$ \citep{mcgee09,delucia12}. The overall mass accretion rate is shown to impact the measured cluster density profiles. In particular, high accretion rates translate in to steep slopes of the density profiles, even in projection at radii $ r> 0.5\,R_{\rm{200,m}}$ \citep{diemer14}. In addition, fast accretion impacts clusters across their entire extent, inducing rapid growth in both $ r_{s}$ and $ r_{\rm{500}}$, with respect to quieter clusters whose inner regions remain undisturbed while still accreting mass in their outskirts \citep{mostoghiu19}. Confirmation of this ongoing accretion is given from the increasing efforts of observations to census infalling haloes at their first encounter with the cluster potential, such as ram-pressure stripped galaxies \citep{poggianti16} and groups \citep{eckert14}. Therefore, a systematic cluster campaign targeting infalling haloes enables simultaneously to address galaxy evolution, cluster physics and physics of hierarchical assembly (e.g. \citealt{haines17, bianconi18}). An observational study of a well-defined cluster sample that would enable direct comparison with theoretical predictions of the impact of merger history on the splashback feature has not been attempted so far.

\begin{deluxetable*}{lcccccc}
\tablecaption{Guide to the LoCuSS datasets used in this article. The references listed in the final column are as follows: 1.~\cite{martino14}; 2.~\cite{okabe16}; 3.~\cite{haines17}; 4.~\cite{bianconi18}; 5.~\cite{sanderson09}; 6.~\cite{mulroy19}.
\label{tab:datasets}}
\tablehead{
Telescope:     & \emph{XMM-Newton} & \emph{Chandra} & Subaru & MMT & UKIRT \\
Instrument:    & EPIC  & ACIS-I & SuprimeCAM & Hectospec & WFCAM \\
Energy / wavelength: & 0.5-2.4 keV & 0.3-7 keV & $V/i'$-bands & $400{-}900{\rm nm}$ & $J/K$-bands \\
Physical radius probed: & $1.5r_{\rm 200,m}$ & $0.8r_{\rm 200,m}$ & $1.5r_{\rm 200,m}$ & $3r_{\rm 200,m}$ & $2.5r_{\rm 200,m}$\\
Sensitivity: & $\rm 10^{-14}\,erg\,s^{-1}\,cm^{-2}$ & $\rm 10^{-13}\,erg\,s^{-1}\,cm^{-2}$  & $V(5\sigma)=27.5$ & 90\% complete & $J(5\sigma)=23$\vspace{-1mm}\\
                            &  &  & $i'(5\sigma)=26$ & at $K\leq K^\star+1.5$ & $K(5\sigma)=22$}
\startdata
Measurements: & \multispan{5}{\dotfill Datasets used\dotfill} & References\\
~~Density profile &  &  &  &  \checkmark & \checkmark & This paper\\
~~Total cluster mass & \checkmark & \checkmark & \checkmark &             &  & 1, 2 \\
~~Infalling groups & \checkmark &       &            & \checkmark & \checkmark & 3, 4\\
~~Central entropy & \checkmark & \checkmark & & & & 5\\
~~X-ray centroid shift & \checkmark & \checkmark & & & & 1\\
~~X-ray concentration & \checkmark & \checkmark & & & & 1\\
~~Luminosity gap & & & \checkmark & \checkmark & \checkmark & 6\\
~~X-ray / BCG offset & \checkmark & \checkmark &  & \checkmark & \checkmark & 6\\
\noalign{\smallskip}
\enddata
\end{deluxetable*}

In this work, we present the detection of the splashback feature in a sample of massive X-ray selected clusters at redshifts of $0.15<z<0.3$ selected from the Local Cluster Substructure Survey (LoCuSS).  Specifically, we take advantage of the rich multi-wavelength LoCuSS dataset and highly complete spectroscopic follow-up observations from the Arizona Cluster Redshift Survey (ACReS) to detect the feature in stacked projected density profile of the clusters, constrain the de-projected splashback radius of the whole sample, and examine how the splashback radius depends on observables that probe the assembly history of the clusters. The use of spectroscopically-confirmed cluster member galaxies, and dissection of the cluster sample based on their properties, are important new steps that can help to shape future investigations of cluster assembly with upcoming surveys including the Vera Rubin Observatory's Legacy Survey of Space and Time, ESA's \emph{Euclid} satellite, ESO's 4MOST instrument, and the German-Russian \emph{eROSITA} X-ray satellite. 

In Section 2 we define the cluster sample, give an overview of the LoCuSS/ACReS dataset, and discuss the structure of the clusters in our sample.  In Section 3, we describe the details of how we approach the modelling of the data.  Section 4 contains our results, beginning with the empirical detection of the splashback feature in the stacked projected density profile of the clusters, before applying the Bayesian modeling scheme described in Section 3 to infer the splashback radius in 3-dimensions, and finally examining how the results depend on the structure of the clusters.  We close by summarising and discussing our results in Section~5.  We assume cosmology values presented in \citet{planck15}, with $ h = 0.678$, $ H_0= 100\,h\,\rm{km\,s^{-1}\,Mpc^{-1}}$, $ \Omega_m = 0.309$ and  $\Omega_{\Lambda}= 0.691$.

\section{Sample, data and cluster properties}\label{sec:dataset}

The Local Cluster Substructure Survey (LoCuSS, PI: G. P. Smith) is a multiwavelength survey of X-ray selected massive galaxy clusters from the \emph{ROSAT} All Sky Survey \citep{ebeling98,boeringer04}.  We study 20 clusters that are the union of the complete high-$L_X$ sample of 50 clusters \citep{okabe13,martino14,okabe16,smith16,mulroy19,farahi19}, and the galaxy evolution sample of 30 clusters \citep{smith10,haines12,haines13,haines15,haines17,bianconi18}.  Therefore, the sample of 20 clusters considered here are an unbiased sub-sample of the complete high-$L_X$ sample, for which high quality wide-field multi-wavelength data are available.  For reference, the high-$L_X$ sample was selected using the following criteria: $0.15\leq z\leq0.3$, $L_X (0.1-2.4 \rm{keV})/E(z) \geq  4.2\times10^{44}\, \rm{erg \, s^{-1}}$, and $\rm -25<\delta [deg]<+65$, $n_H\leq7\times10^{20}{\rm cm^{-2}}$.  The 20 clusters are listed in Table~\ref{table_clusters}. 

The full wide-field multiwavelength dataset on the sample of 20 clusters spans X-ray to millimetre wavelengths, and includes data from \emph{Chandra}, \emph{XMM-Newton}, GALEX, the Subaru 8.2-m telescope, the Hectospec instrument on the Multiple Mirror Telescope (MMT), UKIRT/WFCAM, the Mayall 4-m telescope, \emph{WISE}, \emph{Spitzer}/MIPS, the PACS and SPIRE instruments on \emph{Herschel}, the Sunyaev Zeldovich Array, and ESA's \emph{Planck} satellite.  We concentrate on the X-ray, optical, and near-infrared data in this article, and summarize the cluster properties derived from these data, and previously published LoCuSS articles that contain detailed information in Table~\ref{tab:datasets}.

Highly complete spectroscopic follow-up of stellar-mass selected candidate galaxy cluster member galaxies out to $\approx 3 r_{\rm{200,m}}$ from the Arizona Cluster Redshift Survey (ACReS\footnote{http://herschel.as.arizona.edu/acres/acres.html}) is central to our analysis and results.  This is because we aim to use spectroscopically confirmed member galaxies as test particles to measure the density profile of the clusters on scales well outside the physical scale on which the splashback feature is expected to be found.  The combined UKIRT/WFCAM and MMT/Hectospec dataset allows us to do this to projected clustercentric radii of 6-7 Mpc. The cluster mass measurements from weak-lensing analysis of Subaru observations reaches a typical scaled radius of $r_{\rm 200,c}$ \citep{okabe16}. This, combined with the mass information available to us from the X-ray analysis, allow us to experiment with different approaches to stacking the cluster luminosity density profiles -- i.e. in physical comoving units, or scaled to a common overdensity radius such as $ r_{\rm 200,m}$.  A visual impression of the quality of the data is given in Figure~\ref{rholk_profiles_scaled}, and full details of density profile construction are described in Section~\ref{sec:results}.

The combined LoCuSS/ACReS dataset also allows us to characterize the structure of clusters.  This is important, because the splashback feature in cluster density profiles is associated with infall of material in to the cluster potential, and the structure of clusters seen at different wavelengths is sensitive to their infall history.  In addition to stacking all of the 20 clusters, we therefore present results in Section~\ref{sec:results} based on sub-dividing the sample in to sub-samples selected based on their structural properties that are motivated by in-fall history.  The main results are based on sub-samples selected on the presence of X-ray detected infalling galaxy groups, which indicate active ongoing accretion of galaxies, based on the analysis of \cite{haines17}. Infalling X-ray groups were identified as extended X-ray sources in our XMM images down to $\approx 2 \times 10^{42}\,{\rm erg \, s^{-1}} (M_{\rm 200} \approx 2.5\times10^{13}\, M_{\odot})$ and confirmed to be infalling in to the clusters by identifying group members within the X-ray emitting region of the group with spectroscopic redshifts consistent with that of the host cluster. The X-ray data typically permit us to identify infalling groups within the radial range $0.35-1.3\, r_{\rm 200}$. In addition we included our own analysis of three clusters not in their sample (A\,586, A\,2485, and ZwCl\,0104.4+0048). We excluded Abell 1689 from the subsample of clusters without infalling groups because of the rich structure of this cluster along the line of sight, which is strongly indicative of ongoing merging activity \citep{miralda-escude95,andersson04, peng09}. In addition, we use proxies for cluster merger history including central entropy, X-ray/BCG offset, X-ray brightness concentration and centroid shift, and luminosity gap, as discussed in Appendix~\ref{sec:cluster_structure}).

\section{Density profile modeling methods}\label{sec:profile_fitting}

Empirical profiles have proven successful in describing the  radial variations of density profiles of dark matter haloes, featuring single or multiple power-law behaviours. Among the latter, Jaffe \citep{jaffe83}, Hernquist \citep{hernquist90} and Navarro-Frenk-White \citep{navarro1996} share the same functional form, differing only in the inner and outer power-law slopes (see \citealt{binney08} for a review). 
Here, we choose to model the inner slope of the density profile with a single power-law prescription \citep{einasto74}, corresponding to the virialized part of the halo, together with an additional power-law describing the infall region. These two components are linked by a transitional term which allows for the radial profile to steepen due to the presence of the splashback feature. Following \citet{diemer14} and using the \texttt{Colossus} toolkit \citep{diemer18}, we adopt the radial density profile $\rho(r)$ accordingly:

\begin{eqnarray}
\label{eq:modelstart}
\rho (r) &=& \rho_{\rm Ein}(r) \times f_{\rm trans}(r) + \rho_{\rm infall}(r), \\[5pt]
\rho_{\rm Ein}(r) &=& \rho_{s} \, {\rm exp}\left\{-\frac{2}{\alpha}\left[\left(\frac{r}{r_{s}}\right)^{\alpha}-1\right]\right\}, \\[5pt]
f_{\rm trans}(r) &=& \left[1+\left(\frac{r}{r_{ t}}\right)^{\beta}\right]^{-\gamma/\beta}, \\[5pt]
\rho_{\rm infall}(r) &=& \frac{\rho_0}{1/\Delta + (r/r_{\rm pivot})^{s_e}}
\label{eq:modelend}
\end{eqnarray}
with $ r_{\rm{pivot}} = 1.5 \,{\rm Mpc}$. The parameters $ \rho_s$ and $\rho_0$ are inner and infall scale density, respectively. The parameter $r_{s}$ is the scale radius of the inner profile and $ r_{t}$ is the transition radius between the inner and infall regimes. The exponent $\alpha$ sets the slope of the inner profile, $\beta$ and $\gamma$ set the shape and depth of the transition term, and $s_e$ sets the slope of the outer term \citep{diemer14}. The transition function helps in reconstructing the density profile around the splashback feature, which gets smoothed by the radial averaging \citep{mansfield17}. The parameter $\Delta$ acts as a maximum cutoff density of the outer term to avoid its spurious contribution at small radii. Overall, implementing such truncation has a negligible impact, and we leave this threshold unconstrained. Hence $\alpha$, $\beta$, $\gamma$, $r_{s}$, $r_{t}$, $\rho_{s}$, $\rho_0$ and $s_{e}$ are the free parameters of the model. We considered also models without transition by setting the parameter $\gamma = 0$. We relate the 2d projected density $\Sigma(R)$ observed from LoCuSS to the 3d density $\rho(r)$ via
\begin{equation}
\Sigma(R) = 2 \int^{\infty}_{R} {\rm d} l \, \frac{\rho(r)r dr} {\sqrt{r^2 + R^2}},
\end{equation}
where $R$ is the 2d projected distance from the cluster center. The splashback radius, $r_{\rm sp}$, is a derived parameter in this model, and represents the minimum of the logarithmic derivative $\rm d\,Log (\rho)/ d\,Log (r)$ of the  density profile.

We perform the estimation of best-fit parameters in the context of Bayesian inference, by computing the posterior distribution 
\begin{equation}
    P(\vec{\theta|d}, m) = \frac{\mathcal{L(\vec{\theta})} P(\vec{\theta}| m)}{\mathcal{Z}},
\end{equation}
where $\mathcal{L}$ is the likelihood, $P(\vec{\theta}| m)$ is the prior, and $\mathcal{Z}$ is the evidence, for a given model $m(x_i|\theta)$  evaluated at the data point $x_i$ using parameter values specified by $\vec{\theta}$. 
We adopt a Gaussian likelihood $\mathcal{L}$ for the data $\vec{d}= (x_i,y_i)$, given $\vec{\theta}$:

\begin{equation}
\log \mathcal{L}(\vec{d|\theta}) = -\frac{1}{2}\sum_i^{N_d}\left(\dfrac{y_i - m(x_i|\theta)}{\sigma_i}\right)^2 + \log \mathcal{N}(N_d,\{\sigma_i\}),
\end{equation}
where $\sigma$ is the uncertainty on the data, and $i$ cycles over the number of data points. 
$\mathcal{N}(N_d,\{\sigma_i\})$ acts as normalisation constant to ensure that the integral of the posterior distributions equals to unity.

\begin{deluxetable}{ll}
\tablecaption{Model parameters and their prior properties\label{table_priors}}
%\tablewidth{120pt}
%\tabletypesize{\scriptsize}
\tablehead{
\colhead{Parameter} & \colhead{Prior}
} 
\startdata
$ \rho_0 \,[\rm 10^3\times M_{\odot} h^2/kpc^3]$ & flat:[0.17, 10]   \\
$ s_e$ & flat:[0.1, 10] \\
$ r_s \, [\rm Mpc/h] $& flat:[0.3, 0.7]  \\
$ \rho_s \,[ \rm 10^4 \times M_{\odot} h^2/kpc^3]$ & flat:[7, 50] \\
$ r_t \, [\rm Mpc/h] $ & flat:[1.0, 6.0] \\
$ \alpha$ & flat:[0.1, 0.8] \\
$ \beta$ & flat:[3.0, 9.0] \\
$ \gamma$ & flat:[2.0, 7.0] \\
\enddata
\end{deluxetable}

In this work, we utilise the nested sampling algorithm to compute the Bayesian evidence $\mathcal{Z}$, following the formalism by \citet{skilling04} implemented in CPNest \citep{cpnest}. Nested sampling advantages include the simultaneous estimates of both evidence and posterior samples, and in being easily parallelizable. By splitting the prior volume over intervals of equal likelihood, the evidence results from simple numerical integration, with the addition of reducing the dimensionality of the problem to 1-d. The algorithm allows to evaluate the posteriors on the model parameters
\begin{equation}
\ln \mathcal{P}(\vec{\theta} |d,m) = \ln \big[ \mathcal{L}(\vec{d} | \vec{m}(\vec{\theta}) ){\rm Pr}(\vec{\theta}| m) \big].
\end{equation}
Nested sampling allows  to perform model selection based on marginalized likelihood. Therefore we can directly  quantify and compare the evidence in the data in favour of models with and without the transition feature. We consider eight free parameters ($\rho_0$, $\rho_s$, $r_{\rm t}$, $r_{\rm s}$, $\alpha$, $\beta$, $\gamma$ and $s_{\rm e}$) from the cluster profile model. 
While the number of free parameters is large relative to the number of data points, our primary intention here is not to extract robust constraints on the model parameters, but rather to use the model fits to smoothly interpolate the data to extract constraints on the density profile. Despite this, we are able to run to convergence all the fitting procedures using flat priors (see Table~\ref{table_priors}), without imposing additional a-priori knowledge on the parameter distributions, and to extract data-driven information.

\section{Analysis and Results}\label{sec:results}
\begin{figure*}
\centering
\begin{tabular}{c c c}
        \textbf{Full cluster sample} & \textbf{Clusters w/o groups} & \textbf{Clusters with groups} \\
        \includegraphics[width=0.3\linewidth, keepaspectratio]{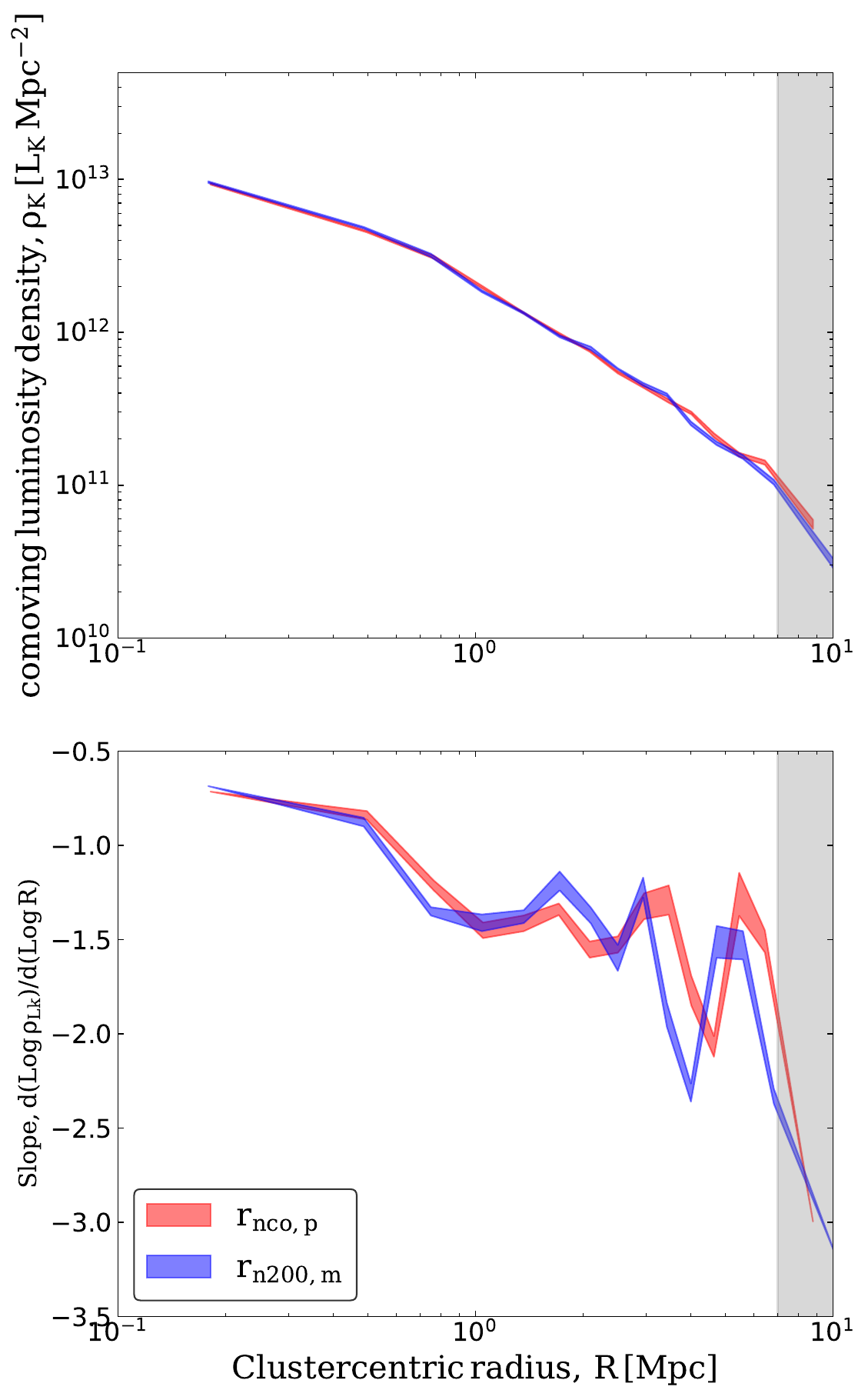} & \includegraphics[width=0.3\linewidth, keepaspectratio]{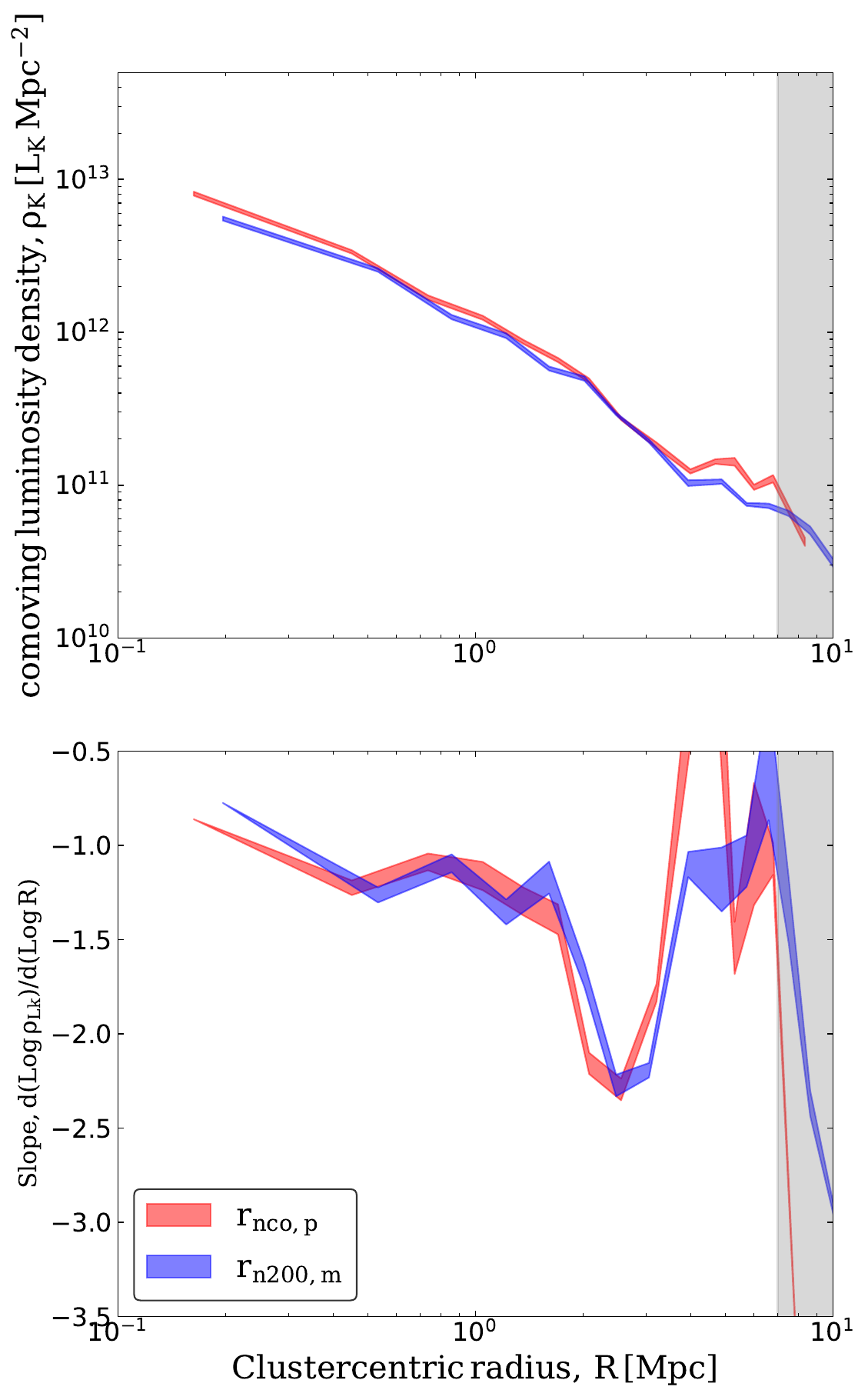} & \includegraphics[width=0.3\linewidth, keepaspectratio]{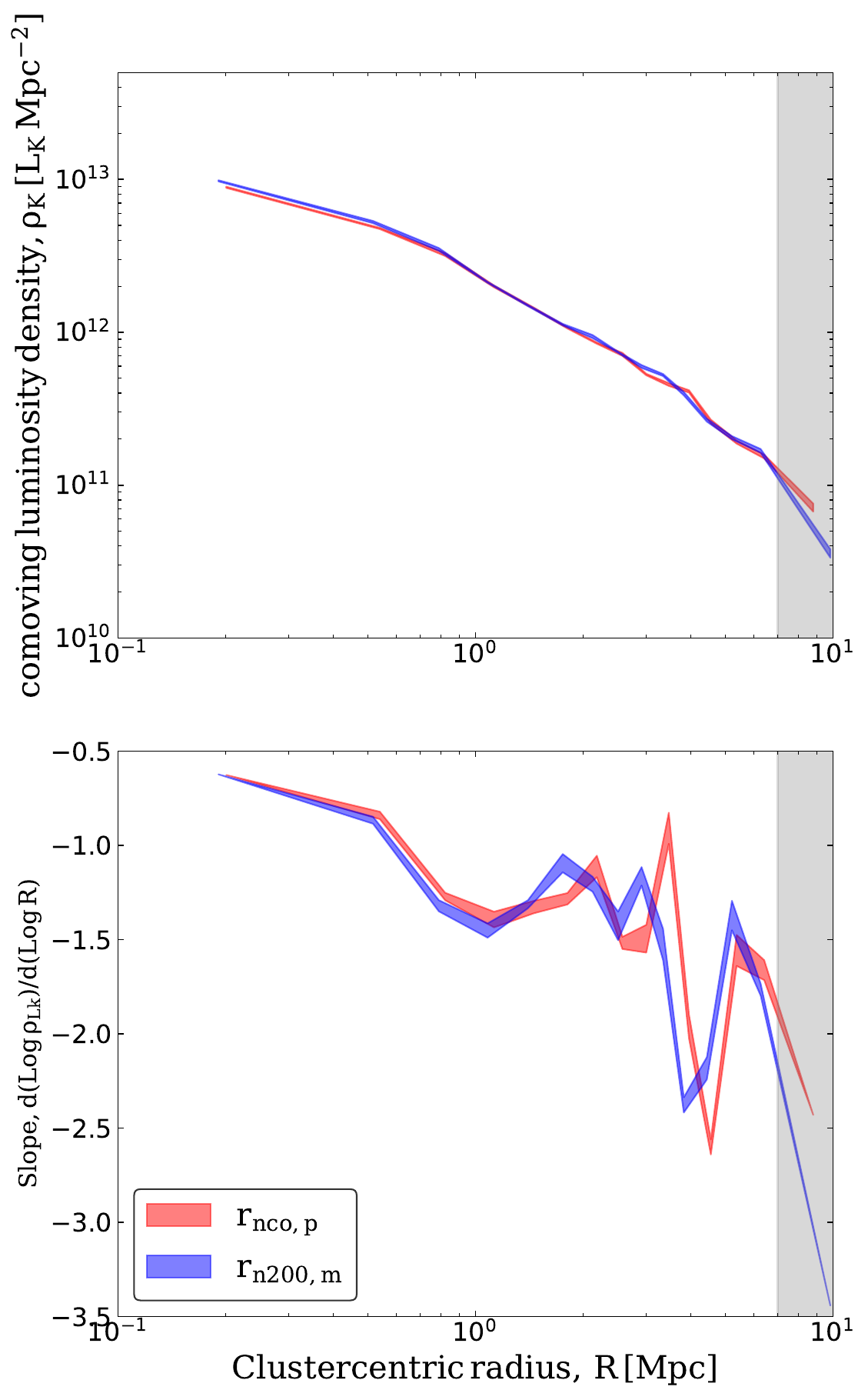}
\end{tabular}
\caption{Top panels: luminosity density profile of the stacked clusters galaxies, computed assigning to each galaxy the mean K-band luminosity of the total sample, using non-scaled and scaled radii according to $ r_{200,m}$ shown in red and blue respectively.
From left to right, we show the profile of the total cluster sample, and the clusters without and with infalling groups.
Bottom panel: logarithmic slope of the luminosity density profile. The colored bands encompasses the errors which are computed from the standard deviation of the mean density within each bin.}\label{plot:nrholk_profiles}
\end{figure*} 

\subsection{Constructing the projected density profile}

Stellar mass, and its close relative near-infrared luminosity is tightly correlated with total cluster mass \citep{Mulroy14,mulroy19}. In addition, \citet{shirasaki21} showed that stellar mass selected galaxy in clusters are good tracers of the gravitational potential of the cluster halo, using the  LoCuSS sample. We therefore use spectroscopically confirmed $K$-band selected cluster galaxies as test particles with which to trace the shape of the cluster density profiles.  We then use the mean $K$-band luminosity of these galaxies and the global cluster $K$-band mass-to-light ratio to convert our number density profiles in to mass density profiles ready for model fitting using the scheme described in Section~3. As a first step, we select spectroscopically confirmed cluster members galaxies down to K-band magnitude $ M_K^*+1.5$, and use them to compute the stacked clustercentric number density profile, centered on position of the brightest cluster galaxy (BCG), which we adopt as the deepest point of each cluster's potential well.  In addition, we  assign the mean K-band luminosity of the whole sample to each galaxy. In this way, we blur-out the effects that different evolution histories may have on the individual galaxies, together with removing any clustercentric distance-dependent process, e.g. mass segregation due to gravitational interactions. Therefore, galaxies are treated as test-particles, that trace the gravitational potential well of the clusters.

We have explored a range of radial bin numbers, and converged on 15 equi-numeric bins to optimise the trade-off between sampling and signal-to-noise of the radial profiles, however our results are not sensitive to this choice, with the binning scheme mainly aiding visualization of the data. 
The projected density $ \Sigma(R)$ is then obtained by multiplying the radially-binned k-corrected K-band luminosity by a mass-to-light ratio $ M/L = 100$, that is consistent with the multi-variate LoCuSS cluster scaling relations \citep{mulroy19}, and other studies of intermediate redshift clusters including \citet{muzzin07}. We have verified that varying the choice of the mass-to-light ratio does not impact or alter the results obtained from the fit presented in Section~\ref{sec:profile_fitting}, in particular regarding the detection of the splashback feature. Indeed, the main motivation for converting light to mass is simply to facilitate the fitting of mass density profiles in Section~\ref{sec:profile_fitting}.

Following \citet{haines13}, each galaxy is weighted by the inverse probability of having being observed spectroscopically. Furthermore, we include an additional weight accounting for the fractional coverage from the near-infrared UKIRT footprint of the circular annuli used to produce the radially-averaged density profiles a function of clustercentric distance. This allows us to quantify the spatial and spectroscopic completeness of our data as a function of increasing clustercentric distance. In particular, we choose to restrict our density profiles to a radius cut corresponding to a weight threshold $ \rm w < 2$, which encompasses the area within which more than 50\% of the galaxies assumed in the analysis are detected.  Figure~\ref{rholk_profiles_scaled} allows us to appreciate more clearly the radial completeness of our sample. In particular, we can see a steep decrease occurring in the profiles beyond $ 2.5 \,r_{200,m}$, which is marked by the grey area. This corresponds to the clustercentric distance at which our sample average completeness weight exceeds the $ w < 2$ threshold discussed above. We note a similar occurring when considering a sample of coeval field galaxies drawn from the LoCuSS dataset, selected in the background and foreground of the clusters \citep{haines15}, which confirms that the completeness threshold is a property of the data, and not related to cluster properties. 

In Figure~\ref{rholk_profiles_scaled}, we show different luminosity profiles computed by arranging the stacked galaxy clustercentric radii according to a range of overdensity thresholds, i.e. $ r_{\rm 200,m}$, $ r_{\rm 200/500,c}$, $ r_{\rm vir}$ and $ r_{\rm 500,x}$. In particular, we consider density contrast with respect to the critical ($c$) and matter density ($m$) of the Universe, and with virial radius obtained from weak-lensing and X-ray ($x$) data respectively. The mean value of these radii for the entire cluster sample is shown by the dashed vertical lines. We note an increasingly self-similar behaviour of the luminosity density profiles when scaled using critical and matter density of the Universe, with respect to the profile computed without using any scaling threshold, marked as $ r_{co,p}$ in the figure. Numerical simulations have shown that the outermost density profiles in clusters at $ r>r_{200,m}$ are self-similar when the radii are scaled by $ r_{200,m}$ (or, more generally, by any radius that is defined with respect to the mean density). This self-similarity indicates that radii defined with respect to the mean density are preferred to describe the structure and evolution of the outer profiles. By contrast, the inner density profiles at smaller radii are most self-similar when radii are scaled by $ r_{200,c}$ \citep{diemer14}. In the following analysis, and in particular regarding fitting, we consider profiles scaled according to  $ r_{200,m}$ from \citeauthor{okabe16}'s weak-lensing measurements.

\begin{table*}
\centering
    \begin{tabular}{l |c|c|c}
    Parameter 
    & \multicolumn{3}{c}{Sample} \\
    and Marginalized posterior& Total & without & with\\
    &  & infalling groups &  infalling groups\\\hline     \hline
$ \rho_0 \,[\rm 10^{3} \times M_{\odot} h^2/kpc^3]$ & $ 4.43_{-1.49}^{+1.41}$  & $ 2.72_{-0.88}^{+0.83}$ & $ 4.11_{-2.05}^{+2.01}$
  \\
$s_e$ & $\rm 1.75_{-0.16}^{+0.15}$ & $ 1.61_{-0.16}^{+0.15}$& $ 1.46_{-0.20}^{+0.19}$\\
$ r_s \, [\rm Mpc/h] $& $ 0.48_{-0.01}^{+0.01}$ & $ 0.48_{-0.05}^{+0.05}$ & $ 0.58_{-0.02}^{+0.02}$\\
$ \rho_s \,[ \rm 10^{5} \times M_{\odot} h^2/kpc^3]$ & $ 246.5_{-24.4}^{+22.1}$ & $150.7_{-33.6}^{+30.3}$ & $ 186.3_{-12.3}^{+12.3}$ \\
$ r_t \, [\rm Mpc/h] $ &$ 4.18_{-0.87}^{+1.00}$  & $ 1.81_{-0.15}^{+0.14}$  & $ 2.78_{-0.12}^{+0.11}$\\
$ \alpha$ & $ 0.27_{-0.02}^{+0.02}$ &  $ 0.19_{-0.05}^{+0.05}$  & $ 0.20_{-0.02}^{+0.02}$\\
$ \beta$ & $>6$ & $>6$   & $>6$\\
$ \gamma$ & $>4$ & $>4$  & $>4$\\
\hline     
$ r_{\rm 200,m} \, [\rm Mpc/h]$ & $2.19\pm 0.01$ & $ 1.74\pm 0.01$ & $ 2.35\pm0.01$\\
$ M_{\rm 200,m} \, [\rm 10^{14}\times M_{\odot}/h]$ & $14.11\pm0.12$ &$ 6.84\pm0.13$ & $ 17.64\pm 0.18$ \\
$ r_{\rm sp} \, [\rm Mpc/h]$ & $3.83 \pm 0.75$  & $ 2.01\pm0.12$  & $ 3.04\pm0.15$\\
max(Log slope 3d) & $-3.4_{-0.6}^{+0.7}$ & $-4.2_{-0.2}^{+0.7}$ & $-4.4_{-0.2}^{+0.9}$\\
    \end{tabular}
    \caption{ Parameters of the best-fit model resulting from the fit of the full cluster sample, and classified according to the presence of infalling groups. The errors quoted in the individual parameters are estimated from their posterior distribution and encompass the $\rm15$ and $\rm85\%$ percentile. The table includes also marginalized posteriors, i.e $ r_{\rm 200,m}$, $ M_{\rm 200,m}$, and 3-d splashback radius  $ r_{\rm sp}$ and their $\pm 3\sigma$ intervals. In addition we quote the minimum values of the 3-d logarithmic slope of the density profile.} \label{table:best-fit_params_full}
 \end{table*} 

\subsection{Empirical detection of splashback feature}

Figure~\ref{plot:nrholk_profiles} (top-left panel) shows the luminosity density profile of the total cluster sample. A dip in the density profile can be noted around 4 Mpc. This feature appears more clearly (bottom-left panel) when plotting the logarithmic slope of the profile $ \epsilon = d\,\rm{Log}(\rho)/\it{d}\,\rm{Log}(R)$, and extends between 3 and 5 Mpc, peaking at 4 Mpc where it reaches $ \epsilon= \rm{-2.30}\pm0.06$. This has a significance of $\rm{5.9} \sigma$ with respect to the mean slope value of $ \epsilon = -1.55\pm0.11$ at neighboring radii. The location of this feature in the density profile, and the run of slope with radius that we obtain are typical of what has been predicted from numerical simulations \citep{diemer14}, even when considering the 2-d surface density. In Figure~\ref{plot:nrholk_profiles}, we show also the density profiles of the cluster sample, split according to the presence of infalling X-ray groups in their surroundings. Interestingly, we notice that the sharp splashback feature, which clearly appears in both sub-samples occurs at smaller radii when considering clusters without in-falling X-ray groups. In particular the splashback feature peaks around 2.5 Mpc with a slope of $\epsilon = \rm{-2.23}\pm0.07$ at $\rm 6.5 \sigma$ for clusters with no infalling groups and peaks at 3.8 Mpc a slope of $ \epsilon = \rm{-2.37}\pm0.06$ at $\rm {6.1} \sigma$ for the systems with infalling groups. Furthermore, we notice that the splashback feature appears consistently at the same location, whether considering scaled or non-scaled radii, for clusters without in-falling groups. This is consistent with clusters that not actively accreting groups having more self-similar structure than clusters that are actively accreting.

We note a similar picture when classifying the clusters using the different structural parameters (see Appendix~\ref{sec:cluster_structure}: Figures~\ref{plot:nrholk_profiles_subsamples}~\&~\ref{plot:nrholk_profiles_subsamples2}). Specifically, cluster sub-samples that are discussed in the literature variously as relaxed, undisturbed, or dynamically quiet (e.g. based on low central entropies, small X-ray centroid shift, large luminosity gap) have a more prominent splashback feature, appearing at smaller clustercentric radii than their so-called unrelaxed, disturbed, or dynamically more active cousins.  The consistency of this picture is very striking because the structural parameters used to define the different sub-samples span a wide range of scales, from central entropy on scales of 20kpc through to the presence of X-ray emitting infalling groups at 1-3Mpc.

\begin{figure*}
 \centering
\begin{tabular}{c c c}
        \hspace{0.9cm}\textbf{Full cluster sample} &  \hspace{0.9cm}\textbf{Clusters w/o groups} &  \hspace{0.9cm}\textbf{Clusters with groups} \\
    \includegraphics[width=0.3\linewidth, keepaspectratio]{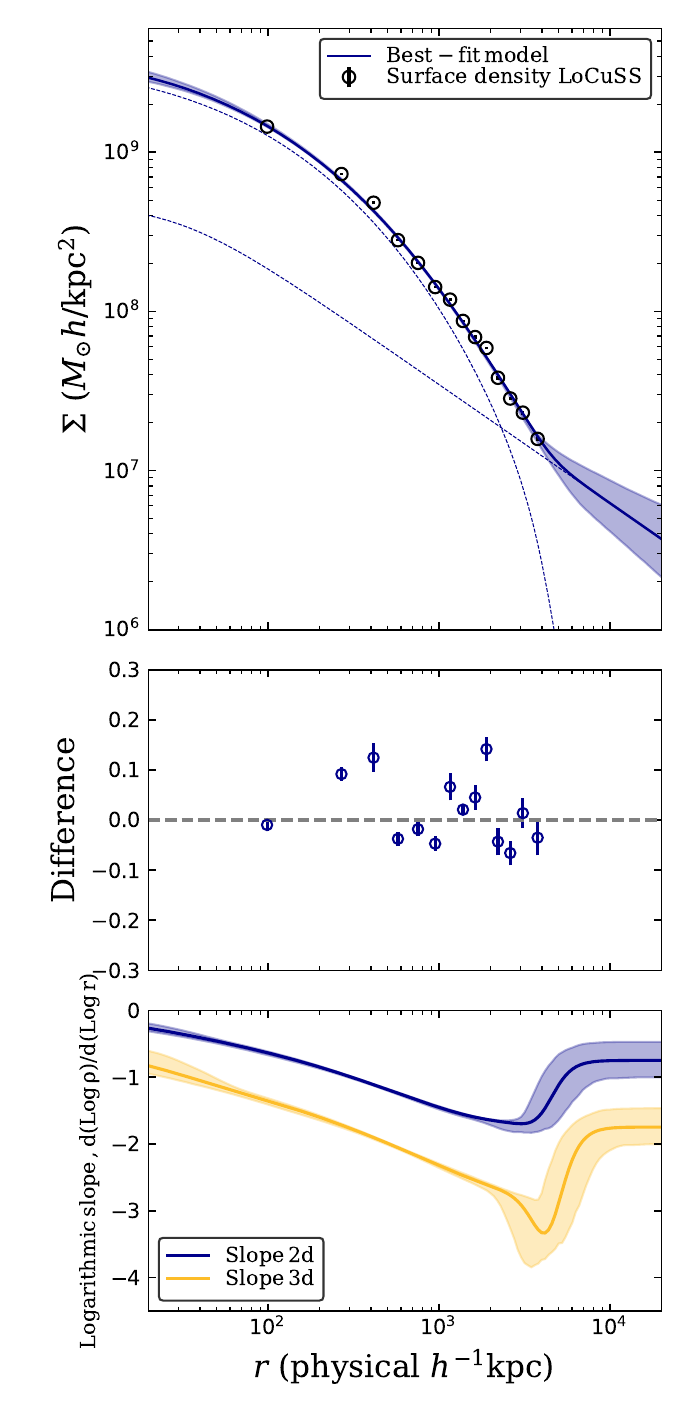} & \includegraphics[width=0.3\linewidth, keepaspectratio]{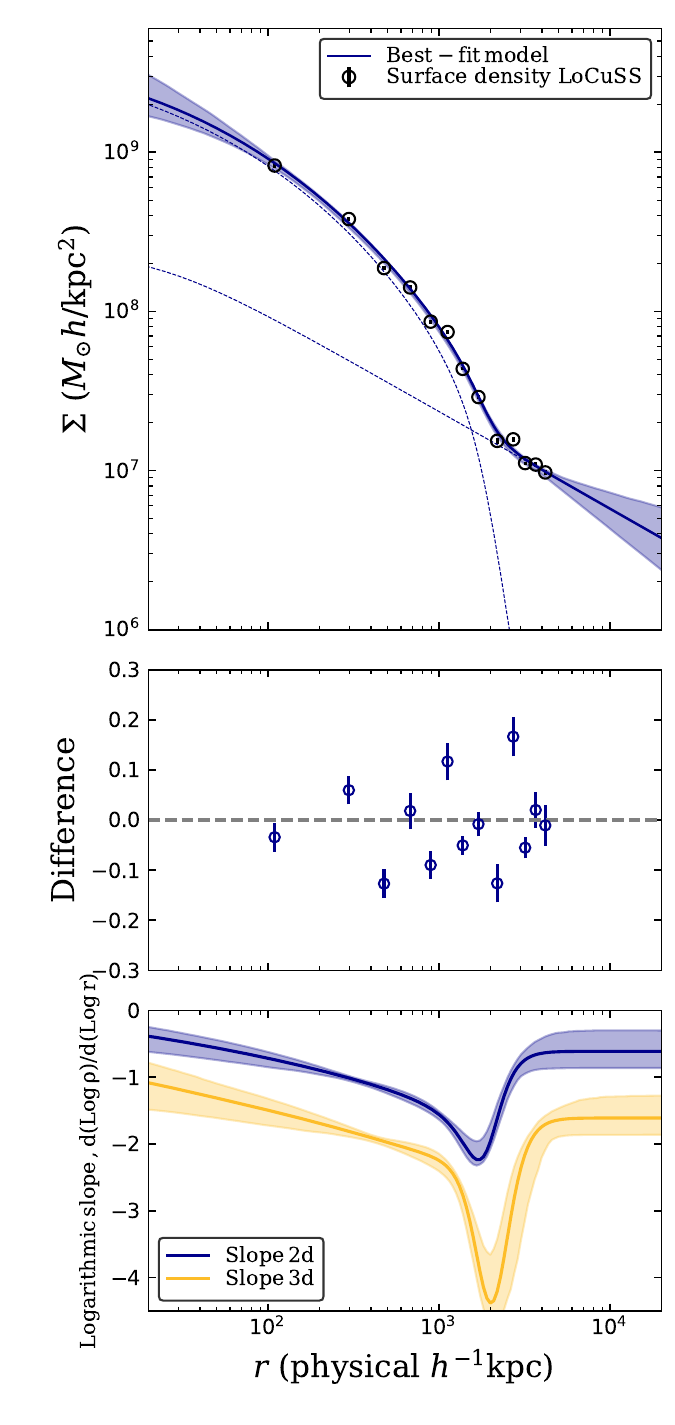} & \includegraphics[width=0.3\linewidth, keepaspectratio]{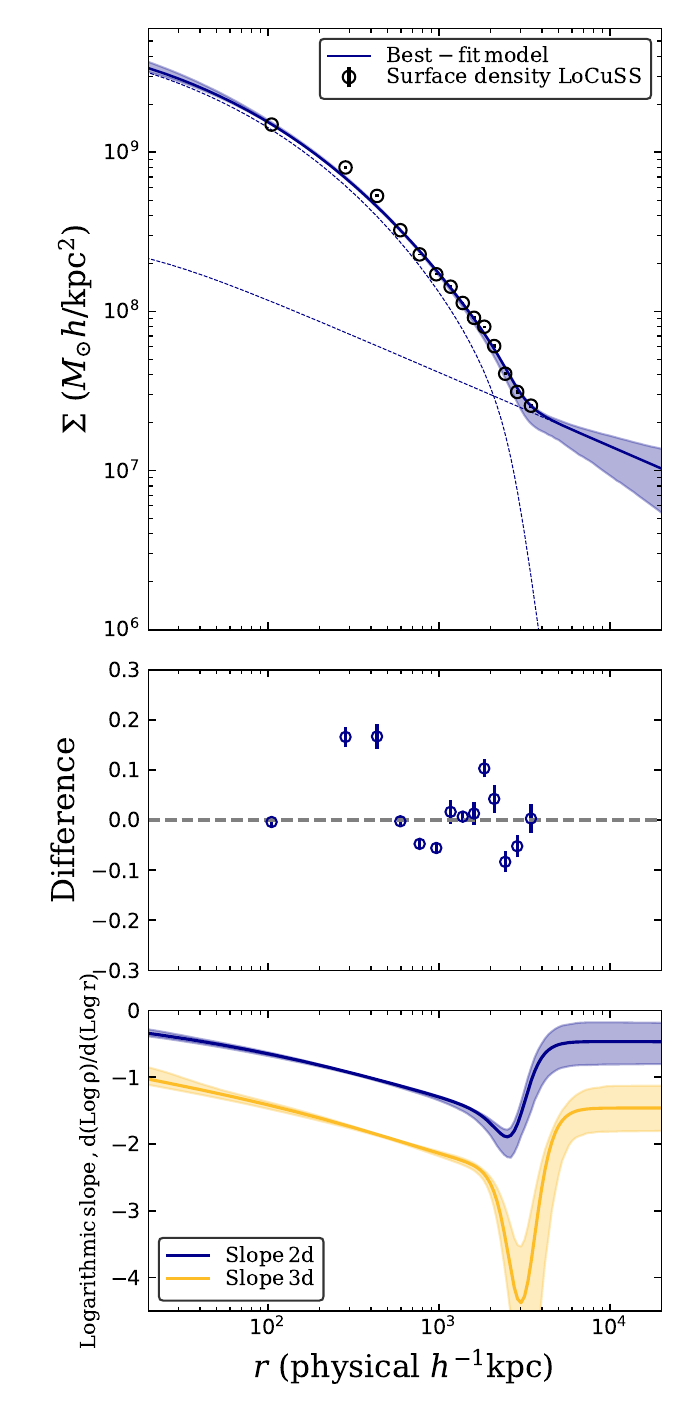}
\end{tabular}    
\caption{Data points and best-fit models for the entire cluster sample (left panel), cluster with no infalling groups (middle panel) and cluster with infalling groups (right panel). For each column, top panel: black circles show the observed surface density profile. Theoretical profile from Section~\ref{sec:profile_fitting} is fitted and plotted as comparison, in dark blue. The profile is split in to its inner and outer components, plotted as dashed lines. Middle panel: difference between the theoretical fitted profiles and data points. Bottom panel: 2D and 3D slopes of the best-fit theoretical model. The shaded area shows the 15 and 85 percentile extracted from the likehood distribution.}\label{plot:data_fit}
\end{figure*}

\subsection{Measurement of splashback radius in 3-dimensions}

We performed the model fit to the entire cluster sample and to sub-samples classified by the presence and absence of infalling groups, as shown in Figure~\ref{plot:data_fit}.  The best-fit parameters, together with salient properties of the best-fit model are summarised in Table \ref{table:best-fit_params_full}.
From the full cluster sample, we recover a ratio between the 3-d splashback radius and $  r_{\rm{200,m}}$ of $ r_{\rm{sp}}/r_{\rm{200,m}} = 1.74 \pm 0.34$ and cluster masses consistent with \citeauthor{okabe16}'s weak-lensing analysis.

As expected based on the results in Section 4.2, the 3-dimensional splashback radius of  clusters without infalling groups, $ r_{\rm{sp}}/r_{\rm{200,m}} = 1.158 \pm 0.071$, is smaller than for clusters with infalling groups $r_{\rm{sp}}/r_{\rm{200,m}} = 1.291 \pm 0.062$.  This difference between the sub-samples is significant at $4.2\sigma$, and persists if we remove the spectroscopically confirmed infalling group members from the cluster member sample. Masses $ M_{\rm{200,m}}$ for both haloes with and without infalling groups are in agreement with the ones recovered from the weak-lensing analysis by \citet{okabe16}. We stress that our dataset is sensitive to the detection of massive infalling groups and retains completeness down to galaxy stellar masses of $ M_{\rm{\star}} = \rm 2\times10^{10}\,M_{\odot}$ \citep{haines17}. Therefore, sampling the full mass range of halo accretion on clusters in beyond the reach of this dataset, hence a direct comparison with simulations remains challenging because predictions that match our observational sample are not yet available. 

Among the parameters considered in the fit, we note looser constraints obtained for $\beta$ and $\gamma$. These parameters are known to be related to the accretion rate of haloes, as shown by numerical simulations \citep{diemer14}, and can jointly span the prior space even in case of haloes with similar properties. This degeneracy is known and has been mitigated by imposing stringent log-normal priors in the literature \citep{shin18, murata20}. We have chosen to not adopt this restriction in our study, after verifying the low impact of the flat priors choice on the fit. In particular, we obtain density profile slopes which are below values of $-3$, which is the lower limit of NFW profiles, when considering the full extent of the posteriors obtained for $\beta$ and $\gamma$. This result provides further evidence of the splashback feature.  The corner plot presenting the distribution of the posteriors of each parameters are shown in Figure~\ref{plot:cornerplot_posteriors_comp} in Appendix~\ref{sec:cluster_structure}.  

The Bayesian framework that we used through the CPNest implementation allowed us to directly compare the models, with and without splashback features, and determine which one is preferred according to the data-driven information. In particular, we can compute directly the Bayes factor $\mathcal{B}$ from the evidence of the two models extracted from the fitting procedure in CPNest, under the assumption of equal and uniform priors. We report a Bayes factor in excess of $\mathcal{B} >100$ in favour of the model with a splashback feature when considering the profiles of clusters without infalling groups. This is the dataset showing the strongest signature of the splashback feature among the ones considered here. The fit to the model without feature outputs a skewed posterior of the $ \rho_0$ parameter towards the lower limit of the prior. This limit corresponds to the critical matter density of the Universe at the redshift considered, and bounds the density domain of the infalling part of the model \citep{diemer14}. This is further evidence of the necessity of a model including a density transition to describe the data profiles. Furthermore, by fitting the same models with the same priors to the data excised of the transition region, we obtain a Bayes factor $\mathcal{B} < 0.6 $, favouring the model not allowing for the splashback transition. 

\section{Summary and discussion}\label{sec:discussion}

\begin{figure*}
 \centering
\includegraphics[width=0.7\linewidth, keepaspectratio]{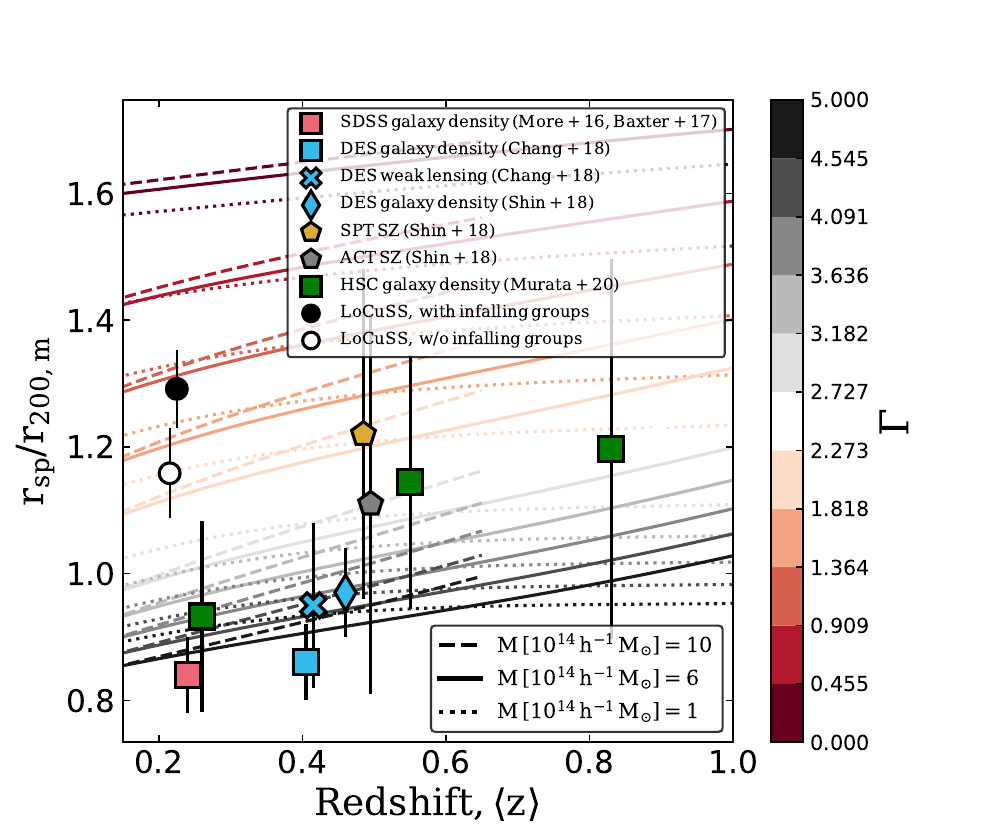}
\caption{Summary of the literature measurements of the 3d splashback feature $ r_{\rm sp}$, normalized by $r_{\rm{200,m}}$, plotted with respect to the mean redshift of the data considered. The shape of the symbol codes the method used for the cluster and galaxy selection, as listed in the corresponding literature reference, and the colour codes the dataset used. LoCuSS datapoint are artificially offset along the x-axis to improve visibility. The background lines mark the redshift evolution of the splashback radius of three reference haloes with masses $ M_{\rm{200,m}} [10^{14} \,h^{-1} \,\rm {M_{\odot}}] \in \{10, 6, 1\}$ as a function of a range of accretion rates, defined as logarithmic mass increment over one dynamical timescale from the empirical relation for cluster-like halo model by \citet{diemer17}.}\label{plot:rsp_comparison}
\end{figure*}

\subsection{Summary of results}

We report the detection of the splashback feature in a sample of massive clusters at intermediate redshifts. The feature is detected using the luminosity density profile of the stacked sample of clusters, computed using the K-band magnitude of spectroscopically confirmed cluster members. Hereafter we list the main results of our analysis:
\begin{itemize}
    \item We empirically detect the splashback feature at a significance greater than $ 5\sigma$. This holds true for the case of the total cluster sample, and for the clusters classified according to the presence/absence of infalling groups. 

    \item We have fitted the observed projected density profiles using the models suggested by \citet{diemer14}, in the context of Bayesian inference in combination with the nested sampling method. This allowed us to recover salient properties of the cluster haloes, including position of the splashback radius relative to $ r_{\rm{200,m}}$. The Bayes factor rates the model allowing for the splashback transition as better describing the data.

    \item The splashback feature position shows a strong dependency according to the presence of infalling groups. Clusters with no detected massive infalling groups present the splashback feature at  $ r_{\rm{sp}}/r_{\rm{200,m}} = 1.158 \pm 0.071$, with respect to cluster accreting groups showing $r_{\rm{sp}}/r_{\rm{200,m}} =  1.291 \pm 0.062$, different at $4.2\sigma$ significance.  This suggests a correlation between the properties of the cluster potential and its accretion rate. We thus report the first measurement of the impact of ongoing accretion and mergers on the measurement of the splashback radius.

    \item Clusters that are classified as old and dynamically inactive present stronger signatures of the splashback feature, with respect to younger, more active clusters. This is not surprising as the latter are bound to show lesser degrees of self-similarity in their density profiles, due to ongoing disturbance caused by accretion and mergers.
    We showed that dynamical properties, despite being defined within the cluster central regions, describe cluster properties which reverberates out to the cluster outskirts.
    
\end{itemize}

\subsection{Comparison with simulations}

 In Figure~\ref{plot:rsp_comparison}, we can see the trend of $r_{\rm{sp}}/r_{\rm{200,m}}$ with respect to redshift. For comparison, the background lines mark the theoretical trend of the ratio $r_{\rm{sp}}/r_{\rm{200,m}}$ as a function of halo accretion rate, which is defined as the logarithmic variation of the cluster mass within one dynamical timescale ($\rm \approx 1 \,Gyr$) from \citet{diemer17}.  Simulations have shown that actively-accreting clusters present a contraction of the splashback radius, correlating with accretion rate. In this work, we observe the opposite behaviour. \citet{sorce20} analysed a set of cluster halos from the MultiDark simulation to find that cluster halos with massive neighbours (with masses above about 10\% of the cluster halos) within $ 2-4\times \, r_{\rm vir}$ had quieter cluster assembly histories recently than on average, and were more active beyond $z\approx 1$.  These halos indeed did not accrete their close-by halos and thus did not empty their neighbourhood. On the contrary, a low number of neighbors in the same distance range is linked to the opposite scenario, namely recently active and quieter in the past. This helps to  reconcile our findings with  the results from simulations by \citet{diemer14}. The clusters showing massive infalling groups are about to enter a phase of substantial accretion, which will result in a contraction of the splashback radius. We nevertheless stress the challenges of capturing the full extent of halo accretion in observed clusters, which is composed of a continuous stream of haloes of widely different masses.  As shown in \citet{diemer14, diemer17} and in Figure~\ref{plot:rsp_comparison}, the ratio $ r_{\rm{sp}}/r_{\rm{200,m}}$ shows the strongest dependence with mass accretion rate. Additionally, at fixed $ r_{\rm{sp}}$, the ratio $ r_{\rm{sp}}/r_{\rm{200,m}}$ depends non-trivially on halo mass, accretion rate and redshift \citep{diemer17}. We note that our results do not show strong dependence on the redshift and halo mass as a result of the narrow interval of both for the cluster sample considered here. Hence, we can further ascribe the observed trend of the ratio $ r_{\rm{sp}}/r_{\rm{200,m}}$ to the accretion rate of clusters. We note that the cluster sample with and without groups are characterised by discrepant mean masses, with the former larger by a factor of $\approx 2.6$. A similar mass separation is found when using the different proxies for cluster dynamical state. However, this is not a major source of systematic uncertainty in our results because the differences in the assembly rate of clusters over the mass and redshift range under consideration are minimal \citep{pizzardo20}. Following the Press and Schechter formalism \citep{lacey93}, the typical mass of infalling haloes is approximately 10\% of the mass of main accreting halo, implying that the typical infalling group mass is correlated with the cluster mass. Therefore, we argue that the detection of infalling groups via X-ray emission in less massive clusters will be impeded by survey limits more than in more massive clusters. This is coupled with the luminosity boost of bright X-ray cluster cores, which impacts favourably the detection of less massive clusters. Crucially, our study confirms that massive clusters are undergoing continuous accretion of group-like haloes, which does not affect the presence of the splashback feature as drastically as for clusters classified via different dynamical proxies (e.g. central entropy). This could be related to the different timescales considered by the different proxies, and with respect to simulations \citep{diemer14,diemer17},  and requires further investigation. Overall, the whole cluster sample provides the less stringent constraints on the recovered splashback feature resulting in a ratio between the 3-d splashback radius and $  r_{\rm{200,m}}$ of $ r_{\rm{sp}}/r_{\rm{200,m}} = 1.74 \pm 0.34$. This supports the scenario in which the dynamical state of individual clusters dilutes the stacked signal of the splashback feature.

\subsection{Comparison with previous observations}

Figure~\ref{plot:rsp_comparison} summarises recently attempted measurements of the splashback feature in cluster samples by means of different observational approaches which include member optical photometric selection (\citealt{more15, baxter17, chang18, shin18, murata20}, see also \citealt{trevisan17}), weak-lensing \citep{chang18}, and SZ \citep{shin18}. Our measurements based on the spectroscopically confirmed cluster galaxies lends further weight to the critical importance of using cluster selection methods based on galaxy membership (cfr. \citealt{shin18}), in terms of both significance and reduction of contamination from interlopers \citep{more15, busch17}. In particular, note that in Figure~\ref{plot:rsp_comparison} we plot $ 3\sigma $ errorbars for data points from our work, and $ \rm 1\sigma $ intervals for measurements from the literature. We stress the  \text{$ 4.2\sigma $}  discrepancy in the ratio $r_{\rm{sp}}/r_{\rm{200,m}}$ between clusters with and without accreting groups. Richness-based methods \citep{more15, busch17, chang18} appear to underestimate the radius at with the splashback transition occurs, partly due to galaxy interloper contamination. Our cluster member selection based on spectroscopic redshifts allows for an efficient removal of field contaminants, and our results are consistent with the model predictions from \citet{murata20}.

\subsection{Future outlook}

Future large-scale surveys, namely \textit{eRosita}, the Vera Rubin Observatory \citep{ivezic19}, and 4MOST \citep{dejong19}, will provide crucial multiwavelength data to sample with greater statistical significance the accretion rates of estimated $\approx 10^5$ clusters, extending the detection of spectroscopically confirmed cluster members and infalling halos at increasing clustercentric distances. Interestingly, \citet{deason21} suggest that the next generation instruments will allow the detection of the splashback feature for the diffuse stellar intracluster light. Understanding the individual impact of halo mass, accretion rate and redshift evolution is currently still an open question, answering which will help towards a more complete description of the fine-grained growth of cosmological structures. Our intent is to promote further discussion between simulations and observations, in particular regarding observational proxies of halo accretion rates.

\section*{Acknowledgements}

MB, RB, GPS, and SLM acknowledge support from the Science and Technology Facilities Council through grant number ST/N021702/1. AB acknowledges support from NSERC (Canada). MB acknowledges Benedikt Diemer for the helpful discussions and support using the Colossus toolkit\footnote{http://www.benediktdiemer.com/code/colossus/}. MB thanks Arya Farahi, August Evrard and Surhud More for the insightful discussions. We warmly thank and acknowledge Maria Pereira and Eiichi Egami for their work on the Arizona Cluster Redshift Survey (ACReS), and for sharing their data with us. Computational work was performed using the University of Birmingham's BlueBEAR HPC service. We acknowledge the use of the Python package \textit{ChainConsumer} \citep{hinton16} for the parameter corner plots.

\bibliographystyle{apj}
\bibliography{biblio.bib}

\begin{thebibliography}{}
\expandafter\ifx\csname natexlab\endcsname\relax\def\natexlab#1{#1}\fi

\bibitem[{{Adhikari} {et~al.}(2014){Adhikari}, {Dalal}, \&
  {Chamberlain}}]{adhikari14}
{Adhikari}, S., {Dalal}, N., \& {Chamberlain}, R.~T. 2014, \jcap, 2014, 019

\bibitem[{{Adhikari} {et~al.}(2020){Adhikari}, {Shin}, {Jain}, {Hilton},
  {Baxter}, {Chang}, {Wechsler}, {Battaglia}, {Bond}, {Bocquet}, {DeRose},
  {Choi}, {Devlin}, {Dunkley}, {Evrard}, {Ferraro}, {Hill}, {Hughes},
  {Gallardo}, {Lokken}, {MacInnis}, {McMahon}, {Madhavacheril}, {Nati},
  {Newburgh}, {Niemack}, {Page}, {Palmese}, {Partridge}, {Rozo}, {Rykoff},
  {Salatino}, {Schillaci}, {Sehgal}, {Sif{\'o}n}, {To}, {Wollack}, {Wu}, {Xu},
  {Aguena}, {Allam}, {Amon}, {Annis}, {Avila}, {Bacon}, {Bertin}, {Bhargava},
  {Brooks}, {Burke}, {Rosell}, {Carrasco Kind}, {Carretero}, {Castander},
  {Choi}, {Costanzi}, {da Costa}, {De Vicente}, {Desai}, {Diehl}, {Doel},
  {Everett}, {Ferrero}, {Fert{\'e}}, {Flaugher}, {Fosalba}, {Frieman},
  {Garc{\'\i}a-Bellido}, {Gaztanaga}, {Gruen}, {Gruendl}, {Gschwend},
  {Gutierrez}, {Hartley}, {Hinton}, {Hollowood}, {Honscheid}, {James},
  {Jeltema}, {Kuehn}, {Kuropatkin}, {Lahav}, {Lima}, {Maia}, {Marshall},
  {Martini}, {Melchior}, {Menanteau}, {Miquel}, {Morgan}, {Ogando},
  {Paz-Chinch{\'o}n}, {Plazas Malag{\'o}n}, {Sanchez}, {Santiago}, {Scarpine},
  {Serrano}, {Sevilla-Noarbe}, {Smith}, {Soares-Santos}, {Suchyta}, {Swanson},
  {Varga}, {Wilkinson}, {Zhang}, {Austermann}, {Beall}, {Becker}, {Denison},
  {Duff}, {Hilton}, {Hubmayr}, {Ullom}, {Van Lanen}, {Vale}, {DES
  Collaboration}, \& {ACT Collaboration}}]{adhikari20}
{Adhikari}, S., {Shin}, T.-h., {Jain}, B., {et~al.} 2020, arXiv e-prints,
  arXiv:2008.11663

\bibitem[{{Andersson} \& {Madejski}(2004)}]{andersson04}
{Andersson}, K.~E., \& {Madejski}, G.~M. 2004, \apj, 607, 190

\bibitem[{{Babul} {et~al.}(2002){Babul}, {Balogh}, {Lewis}, \&
  {Poole}}]{babul02}
{Babul}, A., {Balogh}, M.~L., {Lewis}, G.~F., \& {Poole}, G.~B. 2002, \mnras,
  330, 329

\bibitem[{{Baxter} {et~al.}(2017){Baxter}, {Chang}, {Jain}, {Adhikari},
  {Dalal}, {Kravtsov}, {More}, {Rozo}, {Rykoff}, \& {Sheth}}]{baxter17}
{Baxter}, E., {Chang}, C., {Jain}, B., {et~al.} 2017, \apj, 841, 18

\bibitem[{{Bertschinger}(1985)}]{bertschinger85}
{Bertschinger}, E. 1985, \apjs, 58, 39

\bibitem[{{Bianconi} {et~al.}(2018){Bianconi}, {Smith}, {Haines}, {McGee},
  {Finoguenov}, \& {Egami}}]{bianconi18}
{Bianconi}, M., {Smith}, G.~P., {Haines}, C.~P., {et~al.} 2018, \mnras, 473,
  L79

\bibitem[{{Bildfell} {et~al.}(2008){Bildfell}, {Hoekstra}, {Babul}, \&
  {Mahdavi}}]{bildfell08}
{Bildfell}, C., {Hoekstra}, H., {Babul}, A., \& {Mahdavi}, A. 2008, \mnras,
  389, 1637

\bibitem[{{Binney} \& {Tremaine}(2008)}]{binney08}
{Binney}, J., \& {Tremaine}, S. 2008, {Galactic Dynamics: Second Edition}

\bibitem[{{B{\"o}hringer} {et~al.}(2004){B{\"o}hringer}, {Schuecker}, {Guzzo},
  {Collins}, {Voges}, {Cruddace}, {Ortiz-Gil}, {Chincarini}, {De Grandi},
  {Edge}, {MacGillivray}, {Neumann}, {Schindler}, \& {Shaver}}]{boeringer04}
{B{\"o}hringer}, H., {Schuecker}, P., {Guzzo}, L., {et~al.} 2004, \aap, 425,
  367

\bibitem[{{Bryan} \& {Norman}(1998)}]{bryan98}
{Bryan}, G.~L., \& {Norman}, M.~L. 1998, \apj, 495, 80

\bibitem[{{Bullock} {et~al.}(2001){Bullock}, {Kolatt}, {Sigad}, {Somerville},
  {Kravtsov}, {Klypin}, {Primack}, \& {Dekel}}]{bullock01}
{Bullock}, J.~S., {Kolatt}, T.~S., {Sigad}, Y., {et~al.} 2001, \mnras, 321, 559

\bibitem[{{Busch} \& {White}(2017)}]{busch17}
{Busch}, P., \& {White}, S. D.~M. 2017, \mnras, 470, 4767

\bibitem[{{Cassano} {et~al.}(2010){Cassano}, {Ettori}, {Giacintucci},
  {Brunetti}, {Markevitch}, {Venturi}, \& {Gitti}}]{cassano10}
{Cassano}, R., {Ettori}, S., {Giacintucci}, S., {et~al.} 2010, \apjl, 721, L82

\bibitem[{{Chang} {et~al.}(2018){Chang}, {Baxter}, {Jain}, {S{\'a}nchez},
  {Adhikari}, {Varga}, {Fang}, {Rozo}, {Rykoff}, {Kravtsov}, {Gruen},
  {Hartley}, {Huff}, {Jarvis}, {Kim}, {Prat}, {MacCrann}, {McClintock},
  {Palmese}, {Rapetti}, {Rollins}, {Samuroff}, {Sheldon}, {Troxel}, {Wechsler},
  {Zhang}, {Zuntz}, {Abbott}, {Abdalla}, {Allam}, {Annis}, {Bechtol},
  {Benoit-L{\'e}vy}, {Bernstein}, {Brooks}, {Buckley-Geer}, {Carnero Rosell},
  {Carrasco Kind}, {Carretero}, {D'Andrea}, {da Costa}, {Davis}, {Desai},
  {Diehl}, {Dietrich}, {Drlica-Wagner}, {Eifler}, {Flaugher}, {Fosalba},
  {Frieman}, {Garc{\'\i}a-Bellido}, {Gaztanaga}, {Gerdes}, {Gruendl},
  {Gschwend}, {Gutierrez}, {Honscheid}, {James}, {Jeltema}, {Krause}, {Kuehn},
  {Lahav}, {Lima}, {March}, {Marshall}, {Martini}, {Melchior}, {Menanteau},
  {Miquel}, {Mohr}, {Nord}, {Ogando}, {Plazas}, {Sanchez}, {Scarpine},
  {Schindler}, {Schubnell}, {Sevilla-Noarbe}, {Smith}, {Smith},
  {Soares-Santos}, {Sobreira}, {Suchyta}, {Swanson}, {Tarle}, {Weller}, \& {DES
  Collaboration}}]{chang18}
{Chang}, C., {Baxter}, E., {Jain}, B., {et~al.} 2018, \apj, 864, 83

\bibitem[{{Cole} \& {Lacey}(1996)}]{cole96}
{Cole}, S., \& {Lacey}, C. 1996, \mnras, 281, 716

\bibitem[{{de Jong} {et~al.}(2019){de Jong}, {Agertz}, {Berbel}, {Aird},
  {Alexander}, {Amarsi}, {Anders}, {Andrae}, {Ansarinejad}, {Ansorge},
  {Antilogus}, {Anwand-Heerwart}, {Arentsen}, {Arnadottir}, {Asplund}, {Auger},
  {Azais}, {Baade}, {Baker}, {Baker}, {Balbinot}, {Baldry}, {Banerji},
  {Barden}, \& {Barklem}}]{dejong19}
{de Jong}, R.~S., {Agertz}, O., {Berbel}, A.~A., {et~al.} 2019, The Messenger,
  175, 3

\bibitem[{{De Lucia} {et~al.}(2012){De Lucia}, {Weinmann}, {Poggianti},
  {Arag{\'o}n-Salamanca}, \& {Zaritsky}}]{delucia12}
{De Lucia}, G., {Weinmann}, S., {Poggianti}, B.~M., {Arag{\'o}n-Salamanca}, A.,
  \& {Zaritsky}, D. 2012, \mnras, 423, 1277

\bibitem[{{Deason} {et~al.}(2021){Deason}, {Oman}, {Fattahi}, {Schaller},
  {Jauzac}, {Zhang}, {Montes}, {Bah{\'e}}, {Dalla Vecchia}, {Kay}, \&
  {Evans}}]{deason21}
{Deason}, A.~J., {Oman}, K.~A., {Fattahi}, A., {et~al.} 2021, \mnras, 500, 4181

\bibitem[{{Diemand} {et~al.}(2007){Diemand}, {Kuhlen}, \& {Madau}}]{diemand07}
{Diemand}, J., {Kuhlen}, M., \& {Madau}, P. 2007, \apj, 667, 859

\bibitem[{{Diemer}(2018)}]{diemer18}
{Diemer}, B. 2018, \apjs, 239, 35

\bibitem[{{Diemer} \& {Kravtsov}(2014)}]{diemer14}
{Diemer}, B., \& {Kravtsov}, A.~V. 2014, \apj, 789, 1

\bibitem[{{Diemer} {et~al.}(2017){Diemer}, {Mansfield}, {Kravtsov}, \&
  {More}}]{diemer17}
{Diemer}, B., {Mansfield}, P., {Kravtsov}, A.~V., \& {More}, S. 2017, \apj,
  843, 140

\bibitem[{{Diemer} {et~al.}(2013){Diemer}, {More}, \& {Kravtsov}}]{diemer13b}
{Diemer}, B., {More}, S., \& {Kravtsov}, A.~V. 2013, \apj, 766, 25

\bibitem[{{Ebeling} {et~al.}(1998){Ebeling}, {Edge}, {Bohringer}, {Allen},
  {Crawford}, {Fabian}, {Voges}, \& {Huchra}}]{ebeling98}
{Ebeling}, H., {Edge}, A.~C., {Bohringer}, H., {et~al.} 1998, \mnras, 301, 881

\bibitem[{{Eckert} {et~al.}(2014){Eckert}, {Molendi}, {Owers}, {Gaspari},
  {Venturi}, {Rudnick}, {Ettori}, {Paltani}, {Gastaldello}, \&
  {Rossetti}}]{eckert14}
{Eckert}, D., {Molendi}, S., {Owers}, M., {et~al.} 2014, \aap, 570, A119

\bibitem[{{Einasto} {et~al.}(1974){Einasto}, {Kaasik}, \& {Saar}}]{einasto74}
{Einasto}, J., {Kaasik}, A., \& {Saar}, E. 1974, \nat, 250, 309

\bibitem[{{Farahi} {et~al.}(2020){Farahi}, {Ho}, \& {Trac}}]{farahi20}
{Farahi}, A., {Ho}, M., \& {Trac}, H. 2020, \mnras, 493, 1361

\bibitem[{{Farahi} {et~al.}(2019){Farahi}, {Mulroy}, {Evrard}, {Smith},
  {Finoguenov}, {Bourdin}, {Carlstrom}, {Haines}, {Marrone}, {Martino},
  {Mazzotta}, {O'Donnell}, \& {Okabe}}]{farahi19}
{Farahi}, A., {Mulroy}, S.~L., {Evrard}, A.~E., {et~al.} 2019, Nature
  Communications, 10, 2504

\bibitem[{{Fillmore} \& {Goldreich}(1984)}]{fillmore84}
{Fillmore}, J.~A., \& {Goldreich}, P. 1984, \apj, 281, 1

\bibitem[{{Gozaliasl} {et~al.}(2014){Gozaliasl}, {Finoguenov}, {Khosroshahi},
  {Mirkazemi}, {Salvato}, {Jassur}, {Erfanianfar}, {Popesso}, {Tanaka},
  {Lerchster}, {Kneib}, {McCracken}, {Mellier}, {Egami}, {Pereira},
  {Brimioulle}, {Erben}, \& {Seitz}}]{gozaliasl14}
{Gozaliasl}, G., {Finoguenov}, A., {Khosroshahi}, H.~G., {et~al.} 2014, \aap,
  566, A140

\bibitem[{{Gunn} \& {Gott}(1972)}]{gunn72}
{Gunn}, J.~E., \& {Gott}, III, J.~R. 1972, ApJ, 176, 1

\bibitem[{{Haines} {et~al.}(2013){Haines}, {Pereira}, \& et~al.}]{haines13}
{Haines}, C.~P., {Pereira}, M.~J., \& et~al., S. 2013, \apj, 775, 126

\bibitem[{{Haines} {et~al.}(2015){Haines}, {Pereira}, \& et~al.}]{haines15}
---. 2015, \apj, 806, 101

\bibitem[{{Haines} {et~al.}(2012){Haines}, {Pereira}, {Sanderson}, {Smith},
  {Egami}, {Babul}, {Edge}, {Finoguenov}, {Moran}, \& {Okabe}}]{haines12}
{Haines}, C.~P., {Pereira}, M.~J., {Sanderson}, A.~J.~R., {et~al.} 2012, ApJ,
  754, 97

\bibitem[{{Haines} {et~al.}(2018){Haines}, {Finoguenov}, {Smith}, {Babul},
  {Egami}, {Mazzotta}, {Okabe}, {Pereira}, {Bianconi}, {McGee}, {Ziparo},
  {Campusano}, \& {Loyola}}]{haines17}
{Haines}, C.~P., {Finoguenov}, A., {Smith}, G.~P., {et~al.} 2018, \mnras, 477,
  4931

\bibitem[{{Hernquist}(1990)}]{hernquist90}
{Hernquist}, L. 1990, \apj, 356, 359

\bibitem[{{Hinton}(2016)}]{hinton16}
{Hinton}, S.~R. 2016, The Journal of Open Source Software, 1, 00045

\bibitem[{{Ivezi{\'c}} {et~al.}(2019){Ivezi{\'c}}, {Kahn}, {Tyson}, {Abel},
  {Acosta}, {Allsman}, {Alonso}, {AlSayyad}, {Anderson}, {Andrew}, {Angel},
  {Angeli}, {Ansari}, {Antilogus}, {Araujo}, {Armstrong}, {Arndt}, {Astier},
  {Aubourg}, {Auza}, {Axelrod}, {Bard}, {Barr}, {Barrau}, {Bartlett}, {Bauer},
  {Bauman}, {Baumont}, {Bechtol}, {Bechtol}, {Becker}, {Becla}, {Beldica},
  {Bellavia}, {Bianco}, {Biswas}, {Blanc}, {Blazek}, {Blandford}, {Bloom},
  {Bogart}, \& {Bond}}]{ivezic19}
{Ivezi{\'c}}, {\v{Z}}., {Kahn}, S.~M., {Tyson}, J.~A., {et~al.} 2019, \apj,
  873, 111

\bibitem[{{Jaffe}(1983)}]{jaffe83}
{Jaffe}, W. 1983, \mnras, 202, 995

\bibitem[{{Lacey} \& {Cole}(1993)}]{lacey93}
{Lacey}, C., \& {Cole}, S. 1993, \mnras, 262, 627

\bibitem[{Mahdavi {et~al.}(2013)Mahdavi, Hoekstra, Babul, Bildfell, Jeltema, \&
  Henry}]{mahdavi13}
Mahdavi, A., Hoekstra, H., Babul, A., {et~al.} 2013, The Astrophysical Journal,
  767, 116

\bibitem[{{Mansfield} {et~al.}(2017){Mansfield}, {Kravtsov}, \&
  {Diemer}}]{mansfield17}
{Mansfield}, P., {Kravtsov}, A.~V., \& {Diemer}, B. 2017, \apj, 841, 34

\bibitem[{{Martino} {et~al.}(2014){Martino}, {Mazzotta}, {Bourdin}, {Smith},
  {Bartalucci}, {Marrone}, {Finoguenov}, \& {Okabe}}]{martino14}
{Martino}, R., {Mazzotta}, P., {Bourdin}, H., {et~al.} 2014, MNRAS, 443, 2342

\bibitem[{{Maughan} {et~al.}(2008){Maughan}, {Jones}, {Pierre}, {Andreon},
  {Birkinshaw}, {Bremer}, {Pacaud}, {Ponman}, {Valtchanov}, \&
  {Willis}}]{maughan08}
{Maughan}, B.~J., {Jones}, L.~R., {Pierre}, M., {et~al.} 2008, \mnras, 387, 998

\bibitem[{{McGee} {et~al.}(2009){McGee}, {Balogh}, {Bower}, {Font}, \&
  {McCarthy}}]{mcgee09}
{McGee}, S.~L., {Balogh}, M.~L., {Bower}, R.~G., {Font}, A.~S., \& {McCarthy},
  I.~G. 2009, \mnras, 400, 937

\bibitem[{{Miralda-Escude} \& {Babul}(1995)}]{miralda-escude95}
{Miralda-Escude}, J., \& {Babul}, A. 1995, \apj, 449, 18

\bibitem[{{More} {et~al.}(2015){More}, {Diemer}, \& {Kravtsov}}]{more15}
{More}, S., {Diemer}, B., \& {Kravtsov}, A.~V. 2015, \apj, 810, 36

\bibitem[{{Mostoghiu} {et~al.}(2019){Mostoghiu}, {Knebe}, {Cui}, {Pearce},
  {Yepes}, {Power}, {Dave}, \& {Arth}}]{mostoghiu19}
{Mostoghiu}, R., {Knebe}, A., {Cui}, W., {et~al.} 2019, \mnras, 483, 3390

\bibitem[{{Mulroy} {et~al.}(2014){Mulroy}, {Smith}, {Haines}, {Marrone},
  {Okabe}, {Pereira}, {Egami}, {Babul}, {Finoguenov}, \& {Martino}}]{Mulroy14}
{Mulroy}, S.~L., {Smith}, G.~P., {Haines}, C.~P., {et~al.} 2014, \mnras, 443,
  3309

\bibitem[{{Mulroy} {et~al.}(2019){Mulroy}, {Farahi}, {Evrard}, {Smith},
  {Finoguenov}, {O'Donnell}, {Marrone}, {Abdulla}, {Bourdin}, {Carlstrom},
  {D{\'e}mocl{\`e}s}, {Haines}, {Martino}, {Mazzotta}, {McGee}, \&
  {Okabe}}]{mulroy19}
{Mulroy}, S.~L., {Farahi}, A., {Evrard}, A.~E., {et~al.} 2019, \mnras, 484, 60

\bibitem[{{Murata} {et~al.}(2020){Murata}, {Sunayama}, {Oguri}, {More},
  {Nishizawa}, {Nishimichi}, \& {Osato}}]{murata20}
{Murata}, R., {Sunayama}, T., {Oguri}, M., {et~al.} 2020, arXiv e-prints,
  arXiv:2001.01160

\bibitem[{{Muzzin} {et~al.}(2007){Muzzin}, {Yee}, {Hall}, \& {Lin}}]{muzzin07}
{Muzzin}, A., {Yee}, H.~K.~C., {Hall}, P.~B., \& {Lin}, H. 2007, \apj, 663, 150

\bibitem[{{Navarro} {et~al.}(1996){Navarro}, {Frenk}, \& {White}}]{navarro1996}
{Navarro}, J.~F., {Frenk}, C.~S., \& {White}, S.~D.~M. 1996, \apj, 462, 563

\bibitem[{{Okabe} \& {Smith}(2016)}]{okabe16}
{Okabe}, N., \& {Smith}, G.~P. 2016, \mnras, 461, 3794

\bibitem[{{Okabe} {et~al.}(2013){Okabe}, {Smith}, {Umetsu}, {Takada}, \&
  {Futamase}}]{okabe13}
{Okabe}, N., {Smith}, G.~P., {Umetsu}, K., {Takada}, M., \& {Futamase}, T.
  2013, \apjl, 769, L35

\bibitem[{Peng {et~al.}(2009)Peng, Andersson, Bautz, \& Garmire}]{peng09}
Peng, E.-H., Andersson, K., Bautz, M.~W., \& Garmire, G.~P. 2009, The
  Astrophysical Journal, 701, 1283

\bibitem[{{Pizzardo} {et~al.}(2020){Pizzardo}, {Di Gioia}, {Diaferio}, {De
  Boni}, {Serra}, {Geller}, {Sohn}, {Rines}, \& {Baldi}}]{pizzardo20}
{Pizzardo}, M., {Di Gioia}, S., {Diaferio}, A., {et~al.} 2020, arXiv e-prints,
  arXiv:2005.11562

\bibitem[{{Planck Collaboration} {et~al.}(2016){Planck Collaboration}, {Ade},
  {Aghanim}, {Arnaud}, {Ashdown}, {Aumont}, {Baccigalupi}, {Banday},
  {Barreiro}, {Bartlett}, {Bartolo}, {Battaner}, {Battye}, {Benabed},
  {Beno{\^\i}t}, {Benoit-L{\'e}vy}, {Bernard}, {Bersanelli}, {Bielewicz},
  {Bock}, {Bonaldi}, {Bonavera}, {Bond}, {Borrill}, {Bouchet}, {Boulanger},
  {Bucher}, {Burigana}, {Butler}, {Calabrese}, {Cardoso}, {Catalano},
  {Challinor}, {Chamballu}, {Chary}, {Chiang}, {Chluba}, {Christensen},
  {Church}, {Clements}, {Colombi}, {Colombo}, {Combet}, {Coulais}, {Crill},
  {Curto}, {Cuttaia}, {Danese}, {Davies}, {Davis}, {de Bernardis}, {de Rosa},
  {de Zotti}, {Delabrouille}, {D{\'e}sert}, {Di Valentino}, {Dickinson},
  {Diego}, {Dolag}, {Dole}, {Donzelli}, {Dor{\'e}}, {Douspis}, {Ducout},
  {Dunkley}, {Dupac}, {Efstathiou}, {Elsner}, {En{\ss}lin}, {Eriksen},
  {Farhang}, {Fergusson}, {Finelli}, {Forni}, {Frailis}, {Fraisse},
  {Franceschi}, {Frejsel}, {Galeotta}, {Galli}, {Ganga}, {Gauthier}, {Gerbino},
  {Ghosh}, {Giard}, {Giraud-H{\'e}raud}, {Giusarma}, {Gjerl{\o}w},
  {Gonz{\'a}lez-Nuevo}, {G{\'o}rski}, {Gratton}, {Gregorio}, {Gruppuso},
  {Gudmundsson}, {Hamann}, {Hansen}, {Hanson}, {Harrison}, {Helou},
  {Henrot-Versill{\'e}}, {Hern{\'a}ndez-Monteagudo}, {Herranz}, {Hildebrand t},
  {Hivon}, {Hobson}, {Holmes}, {Hornstrup}, {Hovest}, {Huang}, {Huffenberger},
  {Hurier}, {Jaffe}, {Jaffe}, {Jones}, {Juvela}, {Keih{\"a}nen}, {Keskitalo},
  {Kisner}, {Kneissl}, {Knoche}, {Knox}, {Kunz}, {Kurki-Suonio}, {Lagache},
  {L{\"a}hteenm{\"a}ki}, {Lamarre}, {Lasenby}, {Lattanzi}, {Lawrence}, {Leahy},
  {Leonardi}, {Lesgourgues}, {Levrier}, {Lewis}, {Liguori}, {Lilje},
  {Linden-V{\o}rnle}, {L{\'o}pez-Caniego}, {Lubin}, {Mac{\'\i}as-P{\'e}rez},
  {Maggio}, {Maino}, {Mandolesi}, {Mangilli}, {Marchini}, {Maris}, {Martin},
  {Martinelli}, {Mart{\'\i}nez-Gonz{\'a}lez}, {Masi}, {Matarrese}, {McGehee},
  {Meinhold}, {Melchiorri}, {Melin}, {Mendes}, {Mennella}, {Migliaccio},
  {Millea}, {Mitra}, {Miville-Desch{\^e}nes}, {Moneti}, {Montier}, {Morgante},
  {Mortlock}, {Moss}, {Munshi}, {Murphy}, {Naselsky}, {Nati}, {Natoli},
  {Netterfield}, {N{\o}rgaard-Nielsen}, {Noviello}, {Novikov}, {Novikov},
  {Oxborrow}, {Paci}, {Pagano}, {Pajot}, {Paladini}, {Paoletti}, {Partridge},
  {Pasian}, {Patanchon}, {Pearson}, {Perdereau}, {Perotto}, {Perrotta},
  {Pettorino}, {Piacentini}, {Piat}, {Pierpaoli}, {Pietrobon}, {Plaszczynski},
  {Pointecouteau}, {Polenta}, {Popa}, {Pratt}, {Pr{\'e}zeau}, {Prunet},
  {Puget}, {Rachen}, {Reach}, {Rebolo}, {Reinecke}, {Remazeilles}, {Renault},
  {Renzi}, {Ristorcelli}, {Rocha}, {Rosset}, {Rossetti}, {Roudier},
  {Rouill{\'e} d'Orfeuil}, {Rowan-Robinson}, {Rubi{\~n}o-Mart{\'\i}n},
  {Rusholme}, {Said}, {Salvatelli}, {Salvati}, {Sandri}, {Santos},
  {Savelainen}, {Savini}, {Scott}, {Seiffert}, {Serra}, {Shellard}, {Spencer},
  {Spinelli}, {Stolyarov}, {Stompor}, {Sudiwala}, {Sunyaev}, {Sutton},
  {Suur-Uski}, {Sygnet}, {Tauber}, {Terenzi}, {Toffolatti}, {Tomasi},
  {Tristram}, {Trombetti}, {Tucci}, {Tuovinen}, {T{\"u}rler}, {Umana},
  {Valenziano}, {Valiviita}, {Van Tent}, {Vielva}, {Villa}, {Wade}, {Wandelt},
  {Wehus}, {White}, {White}, {Wilkinson}, {Yvon}, {Zacchei}, \&
  {Zonca}}]{planck15}
{Planck Collaboration}, {Ade}, P.~A.~R., {Aghanim}, N., {et~al.} 2016, \aap,
  594, A13

\bibitem[{{Poggianti} {et~al.}(2016){Poggianti}, {Fasano}, {Omizzolo},
  {Gullieuszik}, {Bettoni}, {Moretti}, {Paccagnella}, {Jaff{\'e}}, {Vulcani},
  {Fritz}, {Couch}, \& {D'Onofrio}}]{poggianti16}
{Poggianti}, B.~M., {Fasano}, G., {Omizzolo}, A., {et~al.} 2016, \aj, 151, 78

\bibitem[{{Ponman} {et~al.}(1994){Ponman}, {Allan}, {Jones}, {Merrifield},
  {McHardy}, {Lehto}, \& {Luppino}}]{ponman94}
{Ponman}, T.~J., {Allan}, D.~J., {Jones}, L.~R., {et~al.} 1994, \nat, 369, 462

\bibitem[{{Poole} {et~al.}(2006){Poole}, {Fardal}, {Babul}, {McCarthy},
  {Quinn}, \& {Wadsley}}]{poole06}
{Poole}, G.~B., {Fardal}, M.~A., {Babul}, A., {et~al.} 2006, \mnras, 373, 881

\bibitem[{{Prada} {et~al.}(2006){Prada}, {Klypin}, {Simonneau},
  {Betancort-Rijo}, {Patiri}, {Gottl{\"o}ber}, \& {Sanchez-Conde}}]{prada06}
{Prada}, F., {Klypin}, A.~A., {Simonneau}, E., {et~al.} 2006, \apj, 645, 1001

\bibitem[{{Pratt} {et~al.}(2019){Pratt}, {Arnaud}, {Biviano}, {Eckert},
  {Ettori}, {Nagai}, {Okabe}, \& {Reiprich}}]{pratt19}
{Pratt}, G.~W., {Arnaud}, M., {Biviano}, A., {et~al.} 2019, \ssr, 215, 25

\bibitem[{{Reiprich} {et~al.}(2013){Reiprich}, {Basu}, {Ettori}, {Israel},
  {Lovisari}, {Molendi}, {Pointecouteau}, \& {Roncarelli}}]{reiprich13}
{Reiprich}, T.~H., {Basu}, K., {Ettori}, S., {et~al.} 2013, \ssr, 177, 195

\bibitem[{{Rykoff} {et~al.}(2014){Rykoff}, {Rozo}, {Busha}, {Cunha},
  {Finoguenov}, {Evrard}, {Hao}, {Koester}, {Leauthaud}, {Nord}, {Pierre},
  {Reddick}, {Sadibekova}, {Sheldon}, \& {Wechsler}}]{rykoff14}
{Rykoff}, E.~S., {Rozo}, E., {Busha}, M.~T., {et~al.} 2014, \apj, 785, 104

\bibitem[{{Sanderson} {et~al.}(2009){Sanderson}, {Edge}, \&
  {Smith}}]{sanderson09}
{Sanderson}, A. J.~R., {Edge}, A.~C., \& {Smith}, G.~P. 2009, \mnras, 398, 1698

\bibitem[{{Schaller} {et~al.}(2015){Schaller}, {Frenk}, {Bower}, {Theuns},
  {Trayford}, {Crain}, {Furlong}, {Schaye}, {Dalla Vecchia}, \&
  {McCarthy}}]{schaller15}
{Schaller}, M., {Frenk}, C.~S., {Bower}, R.~G., {et~al.} 2015, \mnras, 452, 343

\bibitem[{{Shi}(2016)}]{shi16}
{Shi}, X. 2016, \mnras, 459, 3711

\bibitem[{{Shin} {et~al.}(2019){Shin}, {Adhikari}, {Baxter}, {Chang}, {Jain},
  {Battaglia}, {Bleem}, {Bocquet}, {DeRose}, {Gruen}, {Hilton}, {Kravtsov},
  {McClintock}, {Rozo}, {Rykoff}, {Varga}, {Wechsler}, {Wu}, {Zhang}, {Aiola},
  {Allam}, {Bechtol}, {Benson}, {Bertin}, {Bond}, {Brodwin}, {Brooks},
  {Buckley-Geer}, {Burke}, {Carlstrom}, {Carnero Rosell}, {Carrasco Kind},
  {Carretero}, {Castander}, {Choi}, {Cunha}, {Crawford}, {da Costa}, {De
  Vicente}, {Desai}, {Devlin}, {Dietrich}, {Doel}, {Dunkley}, {Eifler},
  {Evrard}, {Flaugher}, {Fosalba}, {Gallardo}, {Garc{\'\i}a-Bellido},
  {Gaztanaga}, {Gerdes}, {Gralla}, {Gruendl}, {Gschwend}, {Gupta}, {Gutierrez},
  {Hartley}, {Hill}, {Ho}, {Hollowood}, {Honscheid}, {Hoyle}, {Huffenberger},
  {Hughes}, {James}, {Jeltema}, {Kim}, {Krause}, {Kuehn}, {Lahav}, {Lima},
  {Madhavacheril}, {Maia}, {Marshall}, {Maurin}, {McMahon}, {Menanteau},
  {Miller}, {Miquel}, {Mohr}, {Naess}, {Nati}, {Newburgh}, {Niemack}, {Ogando},
  {Page}, {Partridge}, {Patil}, {Plazas}, {Rapetti}, {Reichardt}, {Romer},
  {Sanchez}, {Scarpine}, {Schindler}, {Serrano}, {Smith}, {Smith},
  {Soares-Santos}, {Sobreira}, {Staggs}, {Stark}, {Stein}, {Suchyta},
  {Swanson}, {Tarle}, {Thomas}, {van Engelen}, {Wollack}, \& {Xu}}]{shin18}
{Shin}, T., {Adhikari}, S., {Baxter}, E.~J., {et~al.} 2019, \mnras, 487, 2900

\bibitem[{{Shirasaki} {et~al.}(2021){Shirasaki}, {Egami}, {Miyazaki}, \&
  {Okabe}}]{shirasaki21}
{Shirasaki}, M., {Egami}, E., {Miyazaki}, S., \& {Okabe}, N. 2021, arXiv
  e-prints, arXiv:2101.01342

\bibitem[{{Skilling}(2004)}]{skilling04}
{Skilling}, J. 2004, in American Institute of Physics Conference Series, Vol.
  735, American Institute of Physics Conference Series, ed. R.~{Fischer},
  R.~{Preuss}, \& U.~V. {Toussaint}, 395--405

\bibitem[{{Smith} {et~al.}(2010){Smith}, {Haines}, {Pereira}, {Egami}, {Moran},
  {Hardegree-Ullman}, {Babul}, {Rex}, {Rawle}, {Zhang}, {Finoguenov}, {Okabe},
  {Sanderson}, {Edge}, \& {Takada}}]{smith10}
{Smith}, G.~P., {Haines}, C.~P., {Pereira}, M.~J., {et~al.} 2010, \aap, 518,
  L18

\bibitem[{{Smith} {et~al.}(2016){Smith}, {Mazzotta}, {Okabe}, {Ziparo},
  {Mulroy}, {Babul}, {Finoguenov}, {McCarthy}, {Lieu}, {Bah{\'e}}, {Bourdin},
  {Evrard}, {Futamase}, {Haines}, {Jauzac}, {Marrone}, {Martino}, {May},
  {Taylor}, \& {Umetsu}}]{smith16}
{Smith}, G.~P., {Mazzotta}, P., {Okabe}, N., {et~al.} 2016, \mnras, 456, L74

\bibitem[{{Sorce} {et~al.}(2020){Sorce}, {Gottl{\"o}ber}, \& {Yepes}}]{sorce20}
{Sorce}, J.~G., {Gottl{\"o}ber}, S., \& {Yepes}, G. 2020, \mnras,
  arXiv:2007.03695

\bibitem[{{Springel} {et~al.}(2018){Springel}, {Pakmor}, {Pillepich},
  {Weinberger}, {Nelson}, {Hernquist}, {Vogelsberger}, {Genel}, {Torrey},
  {Marinacci}, \& {Naiman}}]{springel18}
{Springel}, V., {Pakmor}, R., {Pillepich}, A., {et~al.} 2018, \mnras, 475, 676

\bibitem[{{Sunyaev} \& {Zeldovich}(1972)}]{sunyaev72}
{Sunyaev}, R.~A., \& {Zeldovich}, Y.~B. 1972, Comments on Astrophysics and
  Space Physics, 4, 173

\bibitem[{{Tinker} {et~al.}(2008){Tinker}, {Kravtsov}, {Klypin}, {Abazajian},
  {Warren}, {Yepes}, {Gottl{\"o}ber}, \& {Holz}}]{tinker08}
{Tinker}, J., {Kravtsov}, A.~V., {Klypin}, A., {et~al.} 2008, \apj, 688, 709

\bibitem[{{Tomooka} {et~al.}(2020){Tomooka}, {Rozo}, {Wagoner}, {Aung},
  {Nagai}, \& {Safonova}}]{tomooka20}
{Tomooka}, P., {Rozo}, E., {Wagoner}, E.~L., {et~al.} 2020, \mnras,
  arXiv:2003.11555

\bibitem[{{Trevisan} {et~al.}(2017){Trevisan}, {Mamon}, \&
  {Stalder}}]{trevisan17}
{Trevisan}, M., {Mamon}, G.~A., \& {Stalder}, D.~H. 2017, \mnras, 471, L47

\bibitem[{{Umetsu} \& {Diemer}(2017)}]{umetsu17}
{Umetsu}, K., \& {Diemer}, B. 2017, \apj, 836, 231

\bibitem[{{Veitch} {et~al.}(2017){Veitch}, {Del Pozzo}, \& {Cody}}]{cpnest}
{Veitch}, J., {Del Pozzo}, W., \& {Cody}. 2017, {Johnveitch/Cpnest: Beta
  Release}, doi:10.5281/zenodo.322819

\bibitem[{{Voit}(2005)}]{voit2005}
{Voit}, G.~M. 2005, Reviews of Modern Physics, 77, 207

\bibitem[{{Walker} {et~al.}(2019){Walker}, {Simionescu}, {Nagai}, {Okabe},
  {Eckert}, {Mroczkowski}, {Akamatsu}, {Ettori}, \& {Ghirardini}}]{walker19}
{Walker}, S., {Simionescu}, A., {Nagai}, D., {et~al.} 2019, \ssr, 215, 7

\bibitem[{{Zu} {et~al.}(2017){Zu}, {Mandelbaum}, {Simet}, {Rozo}, \&
  {Rykoff}}]{zu17}
{Zu}, Y., {Mandelbaum}, R., {Simet}, M., {Rozo}, E., \& {Rykoff}, E.~S. 2017,
  \mnras, 470, 551

\bibitem[{{Z{\"u}rcher} \& {More}(2019)}]{zuercher19}
{Z{\"u}rcher}, D., \& {More}, S. 2019, \apj, 874, 184

\end{thebibliography}

\appendix

\section{Cluster structure}\label{sec:cluster_structure}

We have explored how classifying clusters using different proxies of their dynamical state tracing the properties of the central regions, reverberates at greater radii on the splashback feature. These proxies include central entropy, X-ray surface brightness morphology, i.e. concentration and centroid-shift, K-band luminosity gap between the two most luminous galaxies, the offset between X-ray peak emission and BCG location. We follow \cite{mulroy19} in using the entropy measurements of \cite{sanderson09} to divide the clusters in to those with stronger cooling based on $ K_0(<20\,{\rm \rm{kpc}})<80\,{\rm keV\,cm^{-2}}$ and those with less strong cooling, i.e. larger values of $K_0$.  The X-ray concentration parameter is defined as the ratio of surface brightness at two characteristic radii, the first encompassing the typical size of cool cores and the second the majority of X-ray emission $ c_{\rm{sb}} = \frac{S_X(<40\,\rm{kpc})}{S_X(<400\,\rm{kpc})}$, to maximise the dichotomy between the surface brightness distribution of cool-core and non cool-core clusters \citep{cassano10}. Here, we utilize the concentration measures from \citet{mulroy19}, who extracted it from the X-ray surface brightness maps from \textit{Chandra/ACIS-I} and \textit{XMM–Newton/EPIC} observations. Similarly, the measure of the position of the X-ray emission centroid in circular apertures of increasing radii has been used to label cluster merger activity. In particular, high standard deviation ($ \langle w [0.01 \, r_{\rm{500,x}}] \rangle  > 1 $) of the centroid peak has been associated with dynamically disturbed clusters \citep{poole06, maughan08, mahdavi13}. Here we use the measures from \citet{martino14}, cautioning about projection effects which could hide ongoing line-of-sight mergers.

The properties of the brightest cluster galaxy (BCG) can help identifying the formation history of the halo in which it is hosted. In particular, the older the halo, the higher the magnitude gap between the BCG and the second most luminous galaxies \citep{ponman94,gozaliasl14, farahi20}. This is due to the action of dynamical friction which facilitates the fall of massive galaxies towards the deep end of the cluster potential where the BCG resides. Subsequent mergers result in a dominant central object surrounded by smaller galaxies. We select a threshold value of the K-band magnitude-gap $ \Delta m_{\rm{k,12}} = 0.5$ \citep{smith10}, dividing the bimodal distribution of the full sample. 

\begin{figure*}
\centering
\includegraphics[width=0.30\linewidth, keepaspectratio]{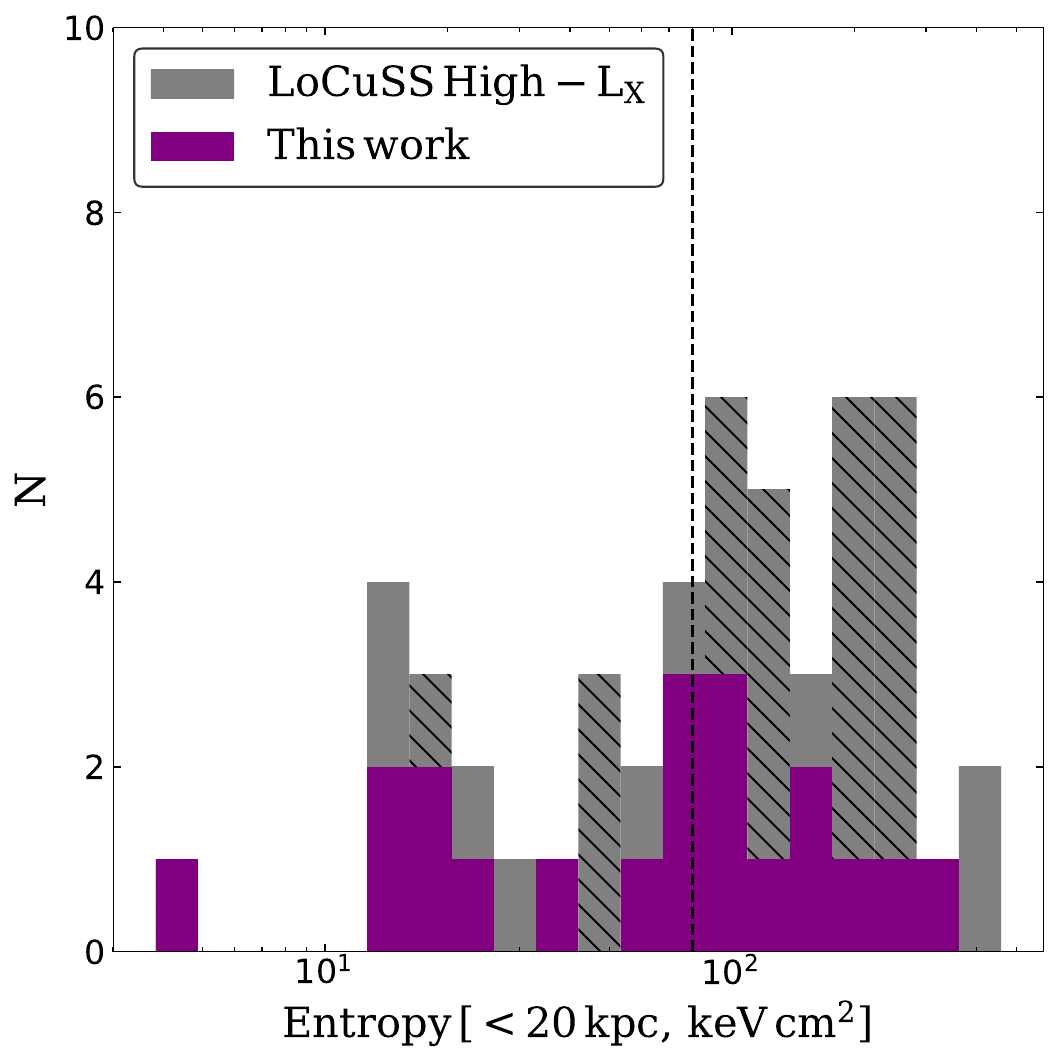}\includegraphics[width=0.30\linewidth, keepaspectratio]{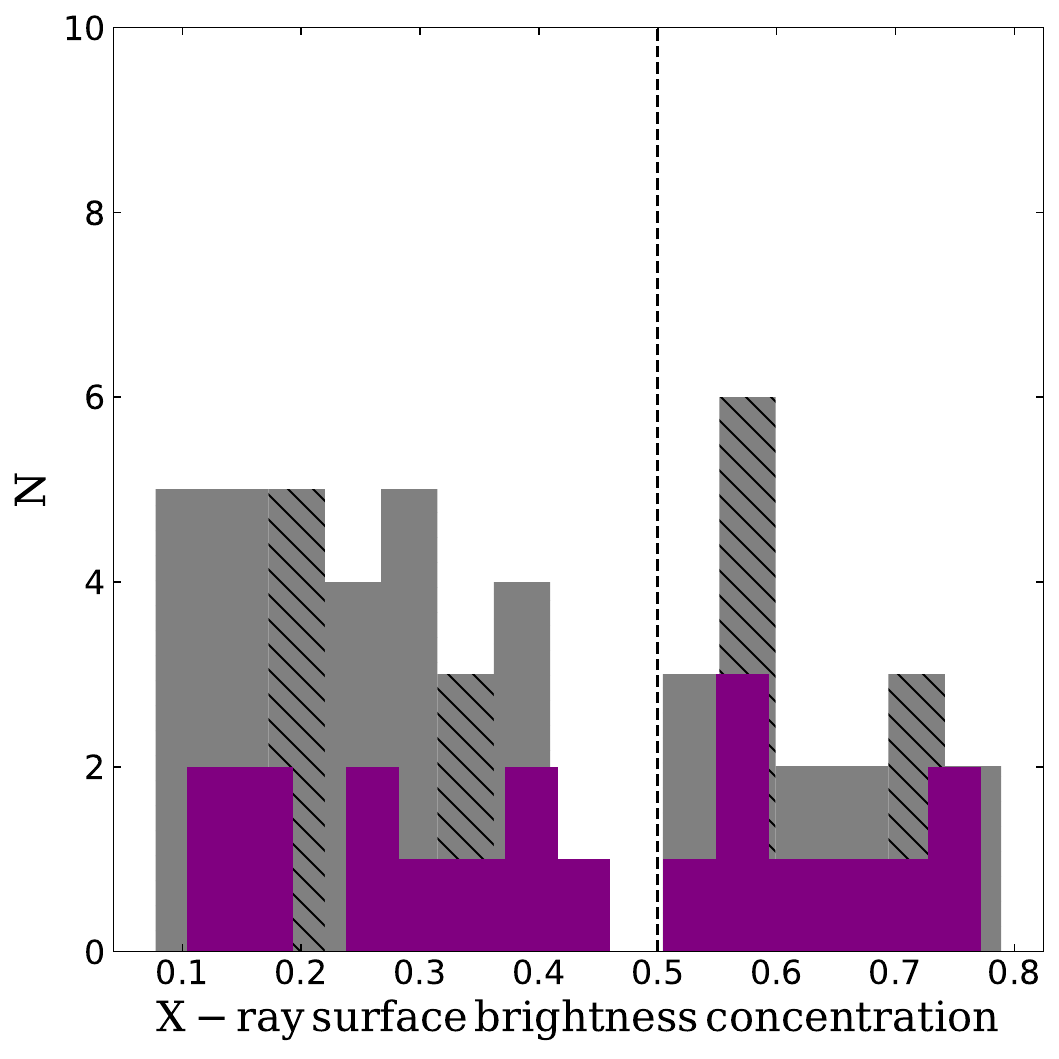}\includegraphics[width=0.30\linewidth, keepaspectratio]{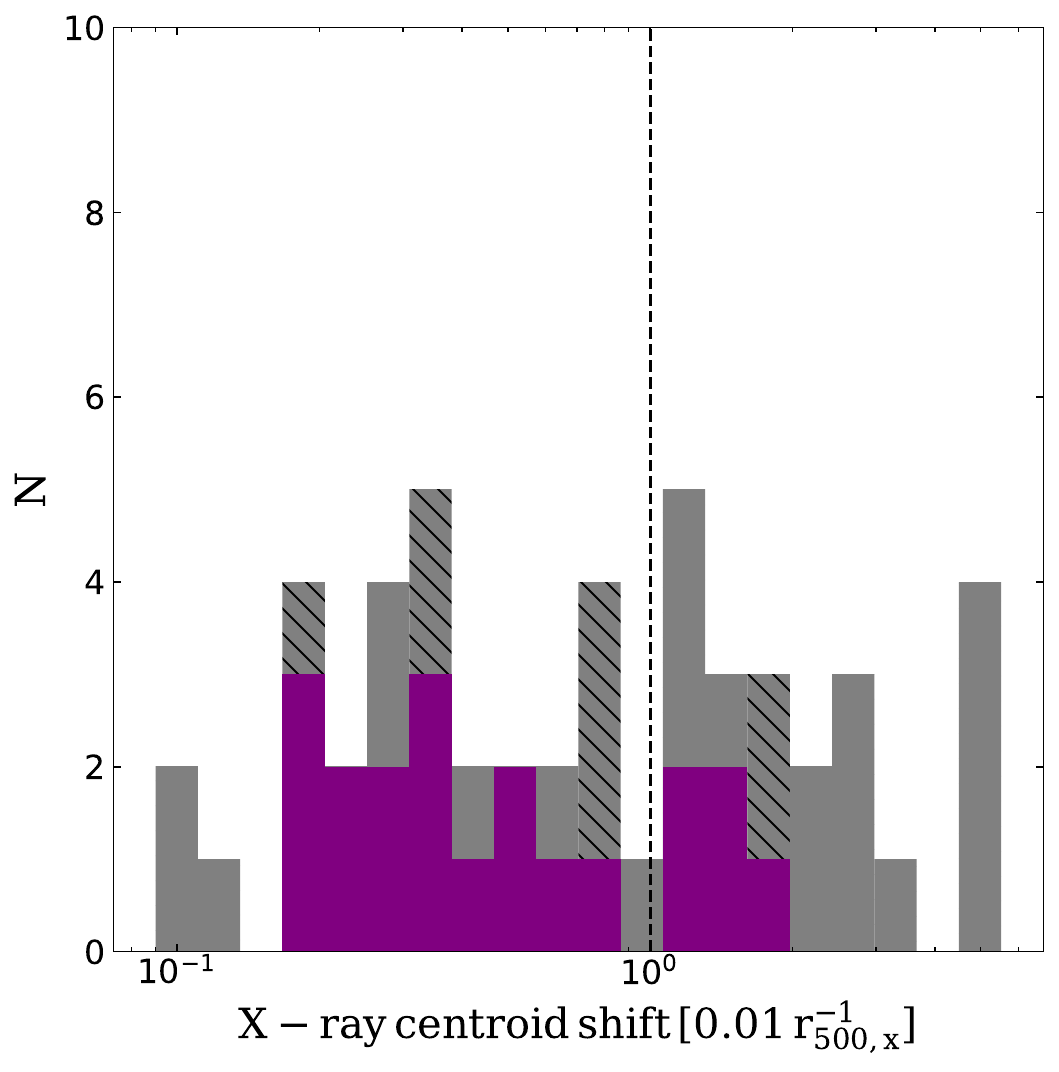}
\includegraphics[width=0.30\linewidth, keepaspectratio]{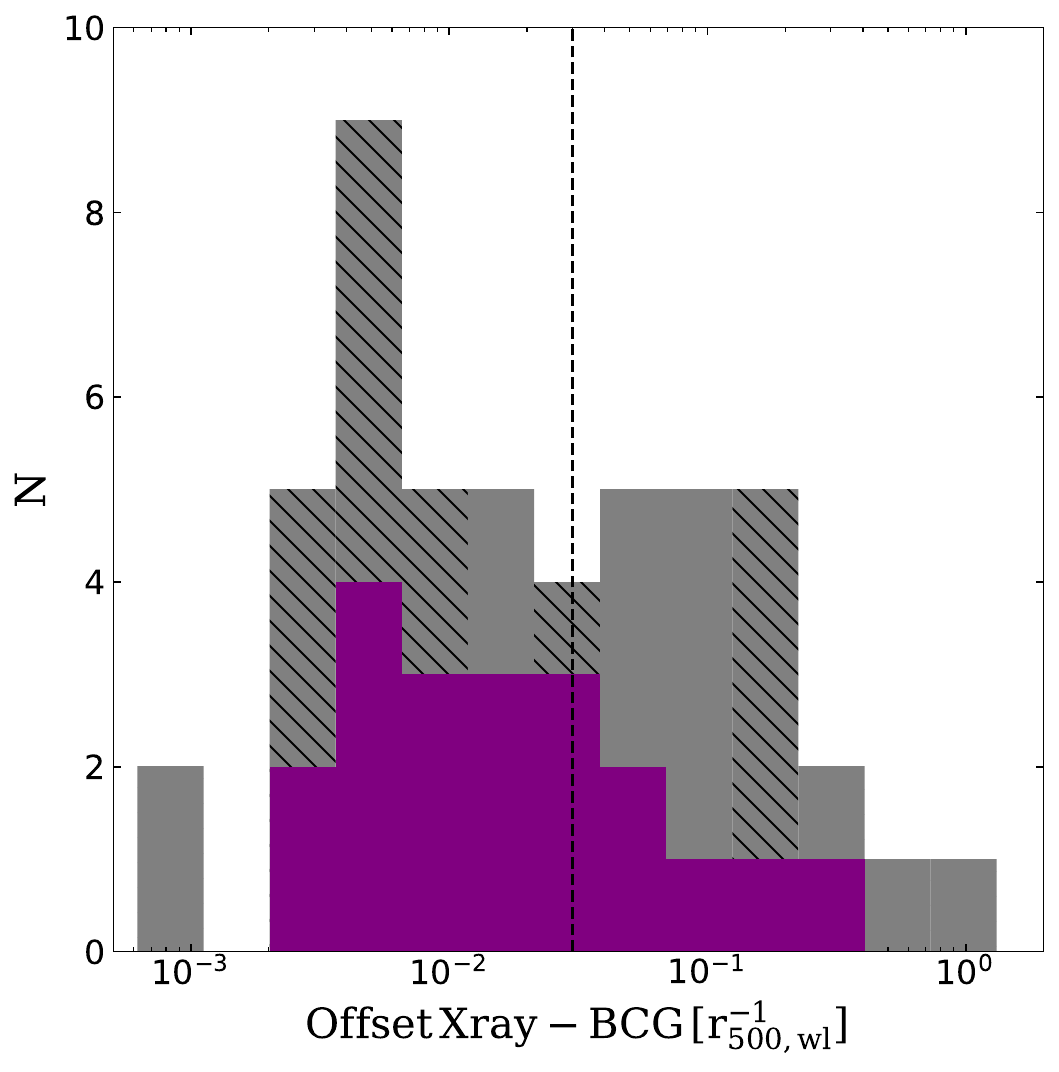}\includegraphics[width=0.30\linewidth, keepaspectratio]{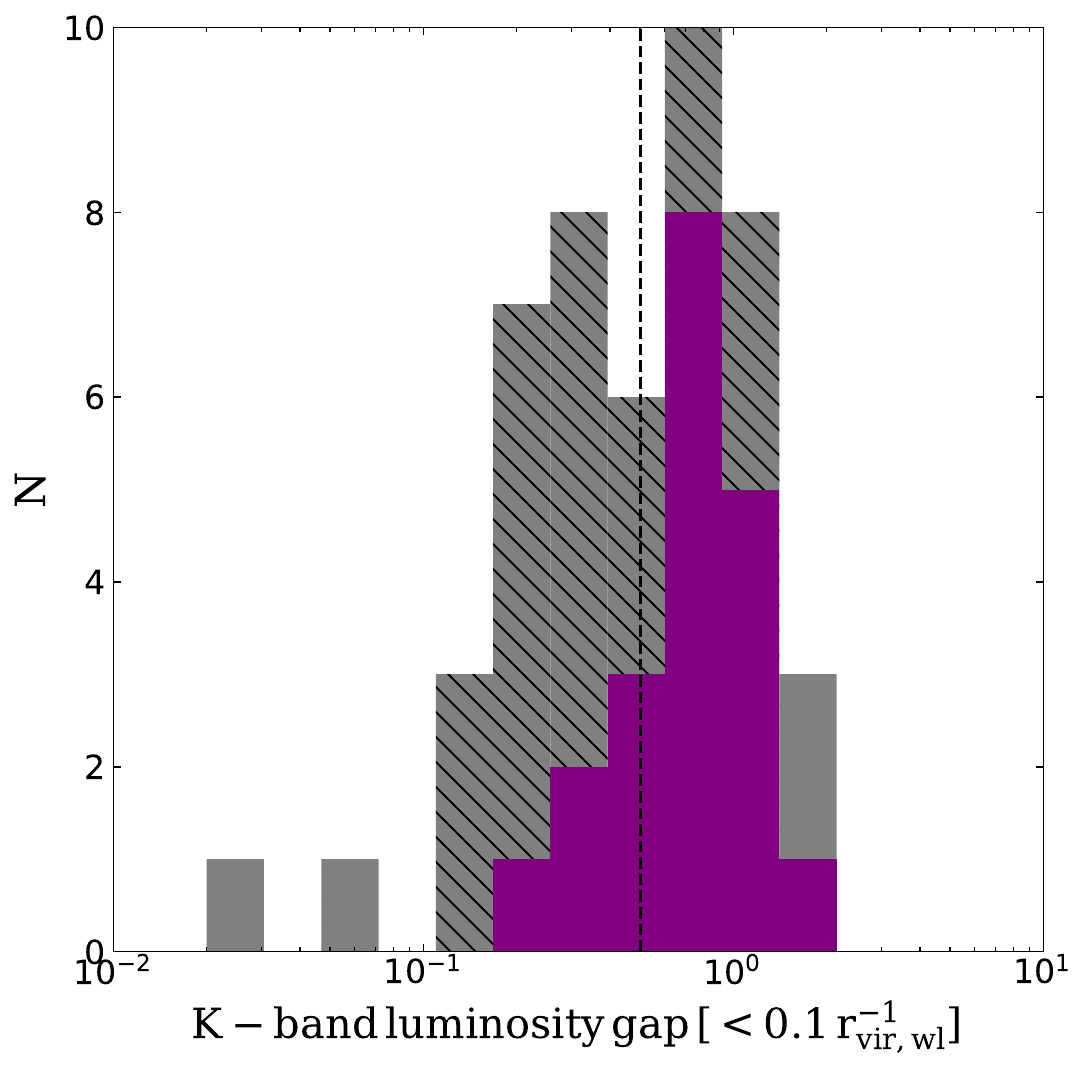}
\caption{Distributions of the salient cluster properties used to classify their dynamical state and formation timescales, as presented in Section~\ref{sec:cluster_structure}. From the top-left to the bottom-right, the panels show the central entropy $ S\,(<20\, \rm{kpc})$, the X-ray surface brightness concentration and centroid shift, the offset between X-ray emission peak and BCG position, and the k-band magnitude gap, respectively. In each panel, the hatched gray and purple histograms show the distribution of the high-$L_X$ LoCuSS sample and subsample presented in Section~\ref{sec:dataset}. The vertical dashed line shows the threshold used to split the sample and produce the plots in Figure~ \ref{plot:nrholk_profiles_subsamples} and \ref{plot:nrholk_profiles_subsamples2} }\label{plot:histo}
\end{figure*}

Bridging between the information provided by the ICM emission and the BCG luminosity, the projected offset between the peak of the X-ray emission and the bulk of the stellar light from the BCG can reveal ongoing cluster dynamical activity \citep{bildfell08}. In particular, in case of old and dynamically-quiet clusters, the X-ray emission and the BCG should coincide indicating the deep core of the cluster potential well.  Following \citet{sanderson09}, we select a threshold value of $ 0.03 \, r_{\rm{500,wl}}$. Figure~\ref{plot:histo} summarises the distributions of the dynamical proxies considered of the parent high-$L_X$ LoCuSS sample, and of the cluster sub-sample used in this work.

\section{Profiles}

\begin{figure*}
 \centering
\includegraphics[width=0.8\linewidth, keepaspectratio]{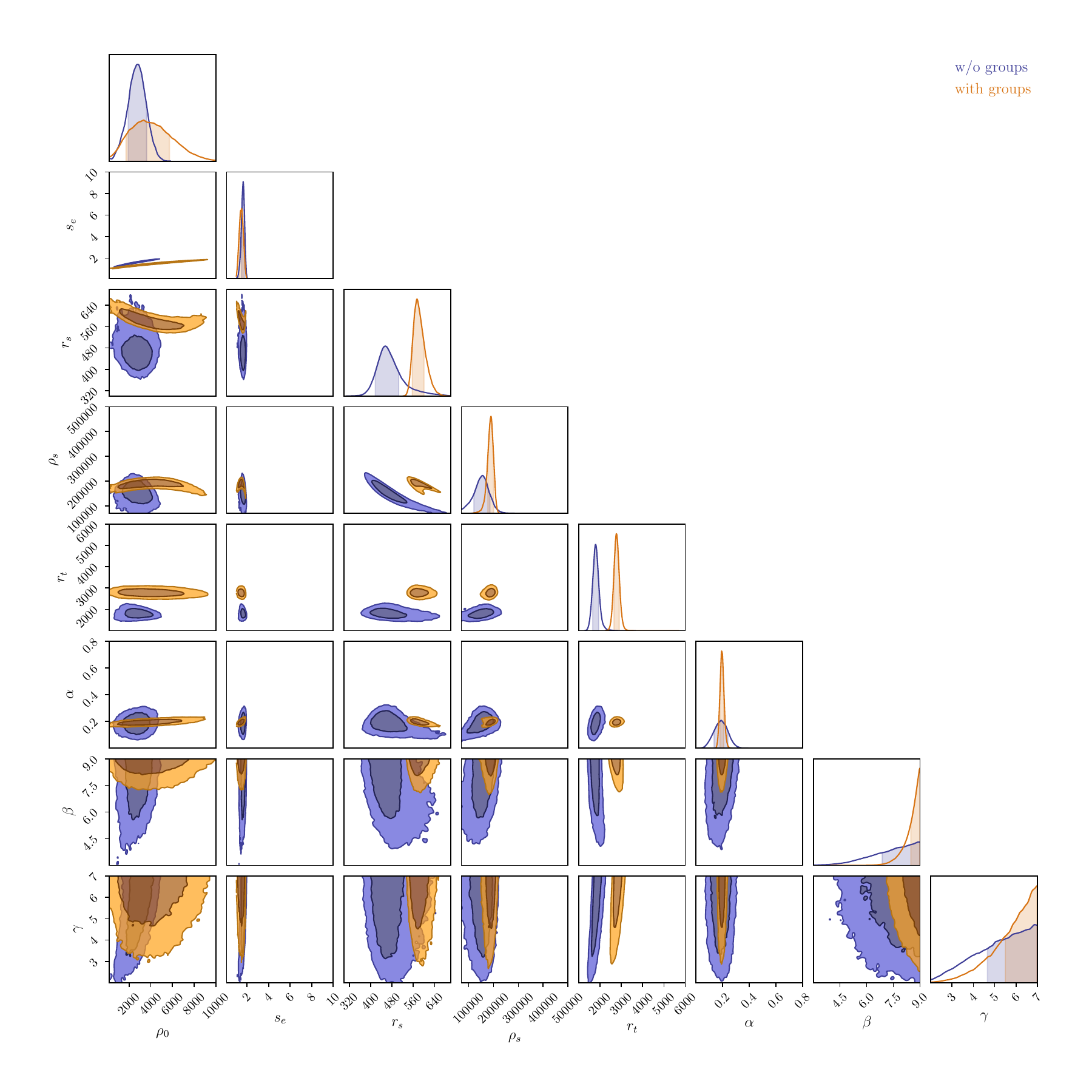}
\caption{Posterior distribution of the parameters of the model fitted to the density profiles of the sample of cluster with and without infalling groups plotted in orange and blue, respectively. Shaded areas in both 1-d and 2-d distributions correspond to 1 $ \sigma$ intervals. The range used in each individual parameter subpanel corresponds to the prior interval. }\label{plot:cornerplot_posteriors_comp}
 \end{figure*}

Figure~ \ref{plot:nrholk_profiles_subsamples} and \ref{plot:nrholk_profiles_subsamples2} show the density profiles of clusters classified according to dynamical proxies as presented in Section~\ref{sec:cluster_structure}. Figure~\ref{plot:cornerplot_posteriors_comp} shows corner plots of the parameters posterior distribution from the fit of the density profiles of clusters with and without infalling groups.

\begin{figure*}
\centering 
\begin{tabular}{c c c c}
        \hspace{0.5cm}\textbf{Low entropy} &  \hspace{0.5cm}\textbf{High entropy} &  \hspace{0.5cm} \hspace{0.5cm}\textbf{Low X-ray centroid shift} &  \hspace{0.5cm}\textbf{High X-ray centroid shift}  \\
\includegraphics[width=0.23\linewidth, keepaspectratio]{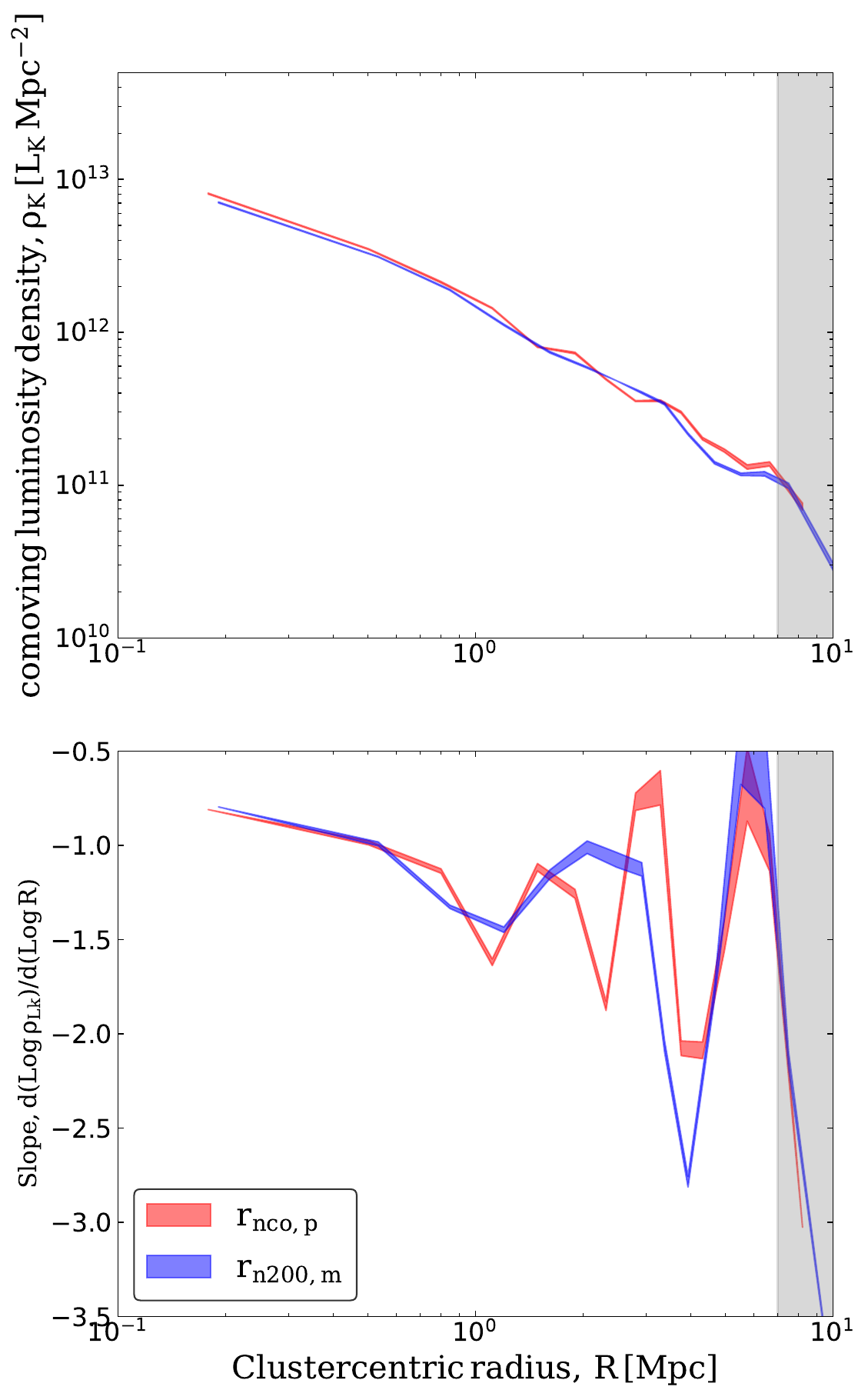} & \includegraphics[width=0.23\linewidth, keepaspectratio]{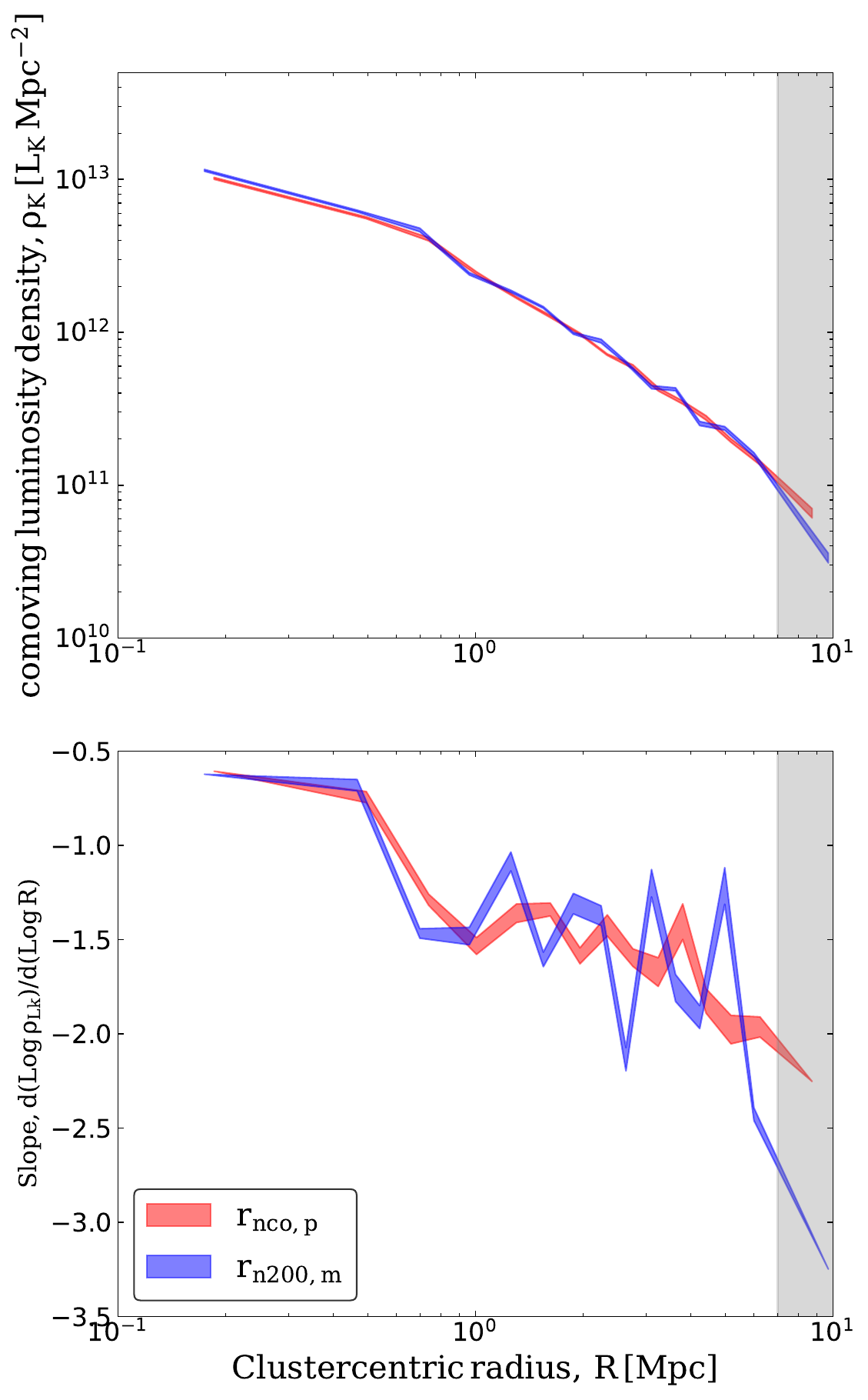} & \includegraphics[width=0.23\linewidth, keepaspectratio]{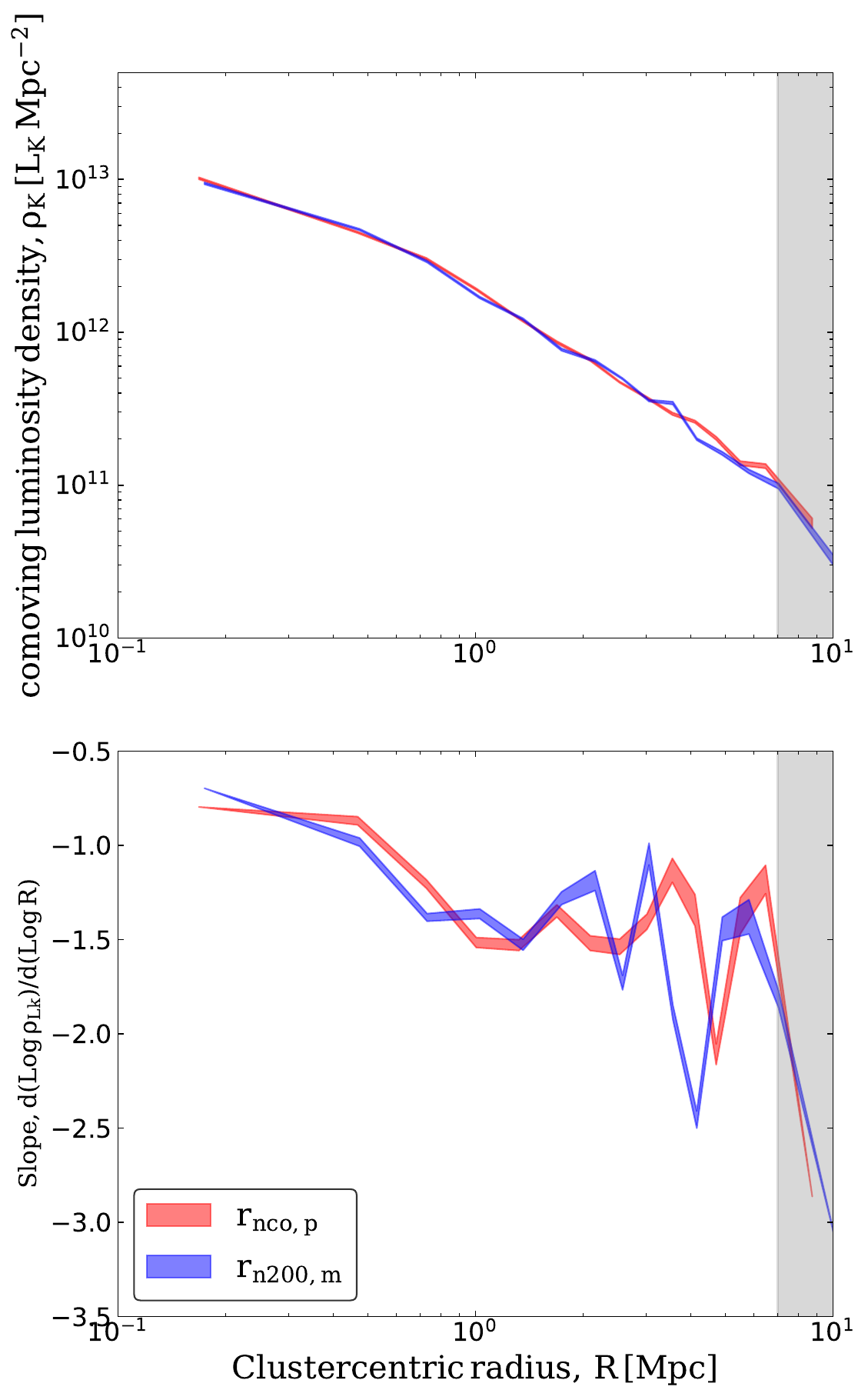} & \includegraphics[width=0.23\linewidth, keepaspectratio]{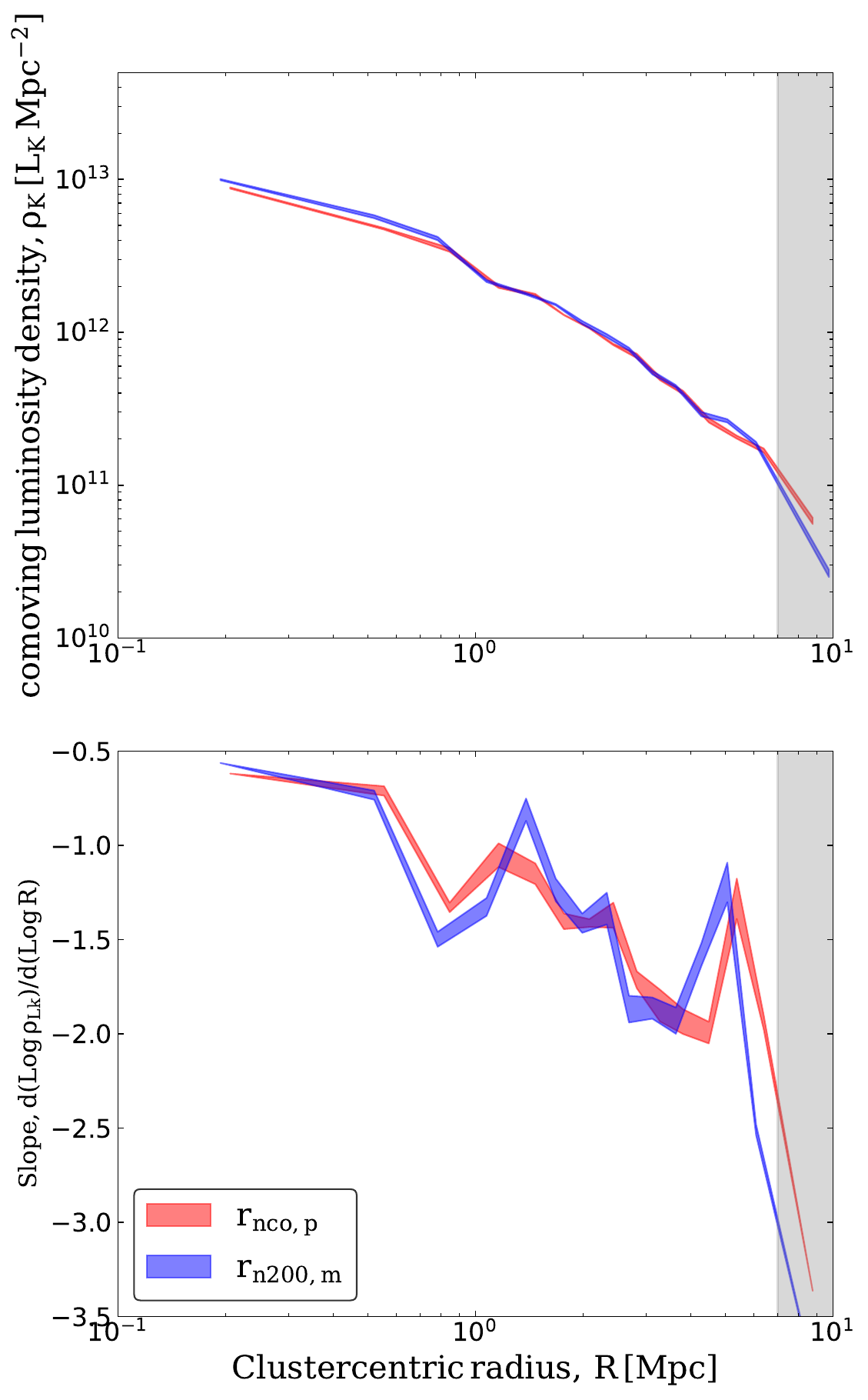}
\end{tabular}

\begin{tabular}{c c}
        \hspace{0.5cm}\textbf{Low X-ray surface brightness concentration} &  \hspace{0.5cm}\textbf{High X-ray surface brightness concentration} \\
\includegraphics[width=0.23\linewidth,keepaspectratio]{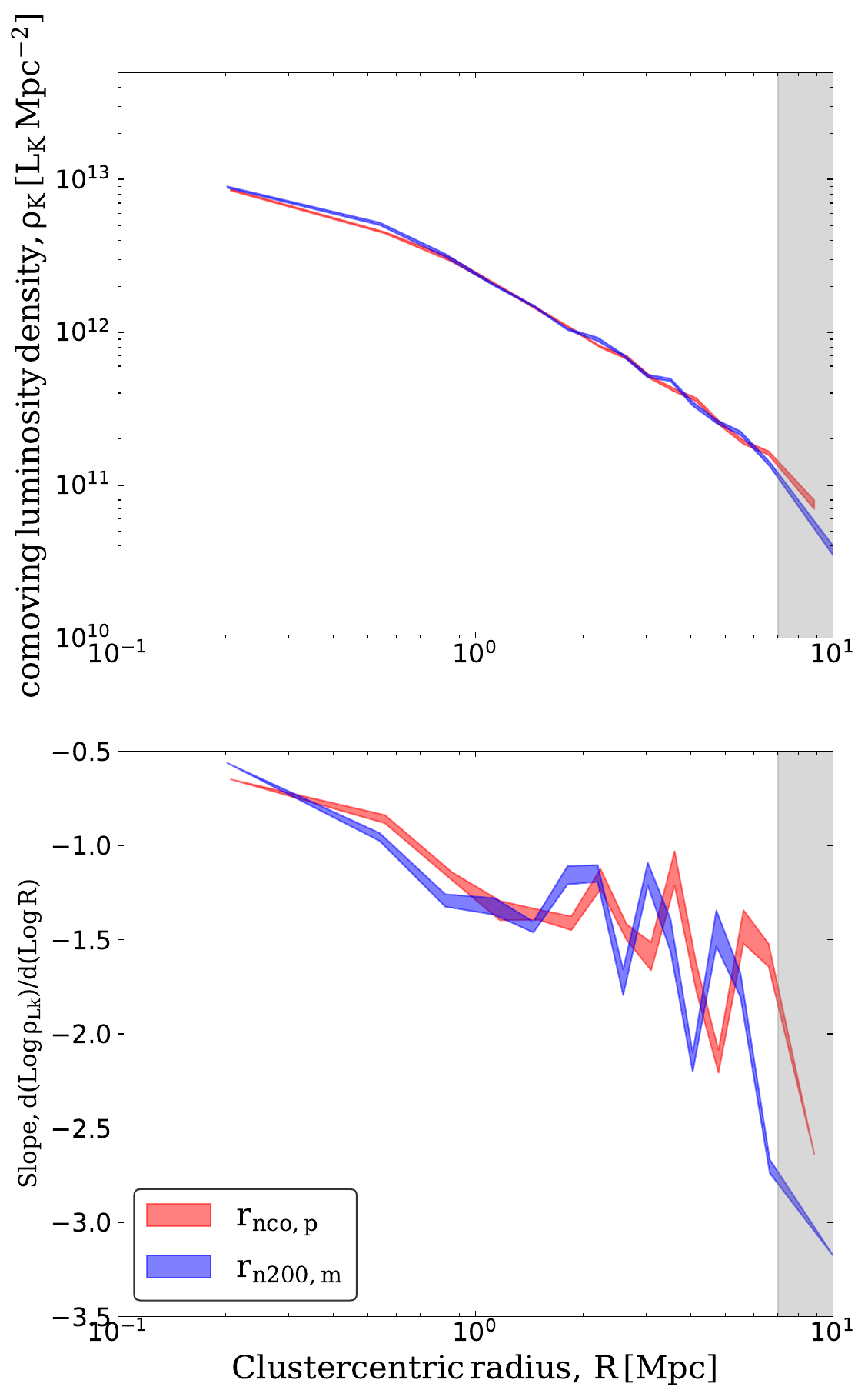} & \includegraphics[width=0.23\linewidth,keepaspectratio]{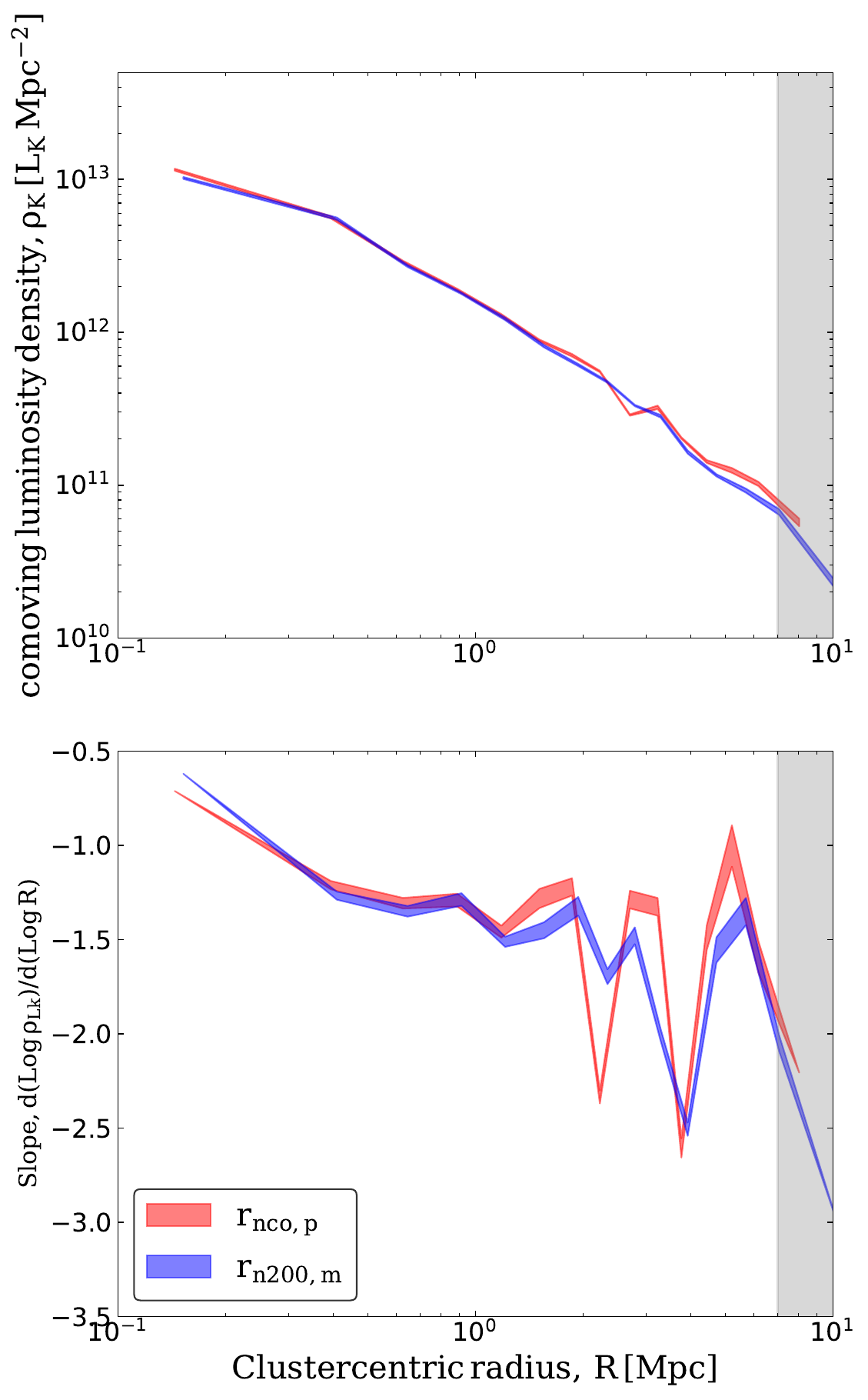}%
\end{tabular}

\caption{In each subpanel, density profiles (top) and slopes (bottom) resulting from the cluster classification according to the different dynamical proxies presented in Section~\ref{sec:cluster_structure}. Continues in Figure~\ref{plot:nrholk_profiles_subsamples2}.}\label{plot:nrholk_profiles_subsamples}
\end{figure*}

\begin{figure*}
\centering 

\begin{tabular}{c c c c}
        \hspace{0.5cm}\textbf{Low K-band luminosity gap} &  \hspace{0.5cm}\textbf{High K-band luminosity gap} &  \hspace{0.5cm} \hspace{0.5cm}\textbf{Low X-ray-BCG offset} &  \hspace{0.5cm}\textbf{High X-ray-BCG offset}  \\
\includegraphics[width=0.25\linewidth,keepaspectratio]{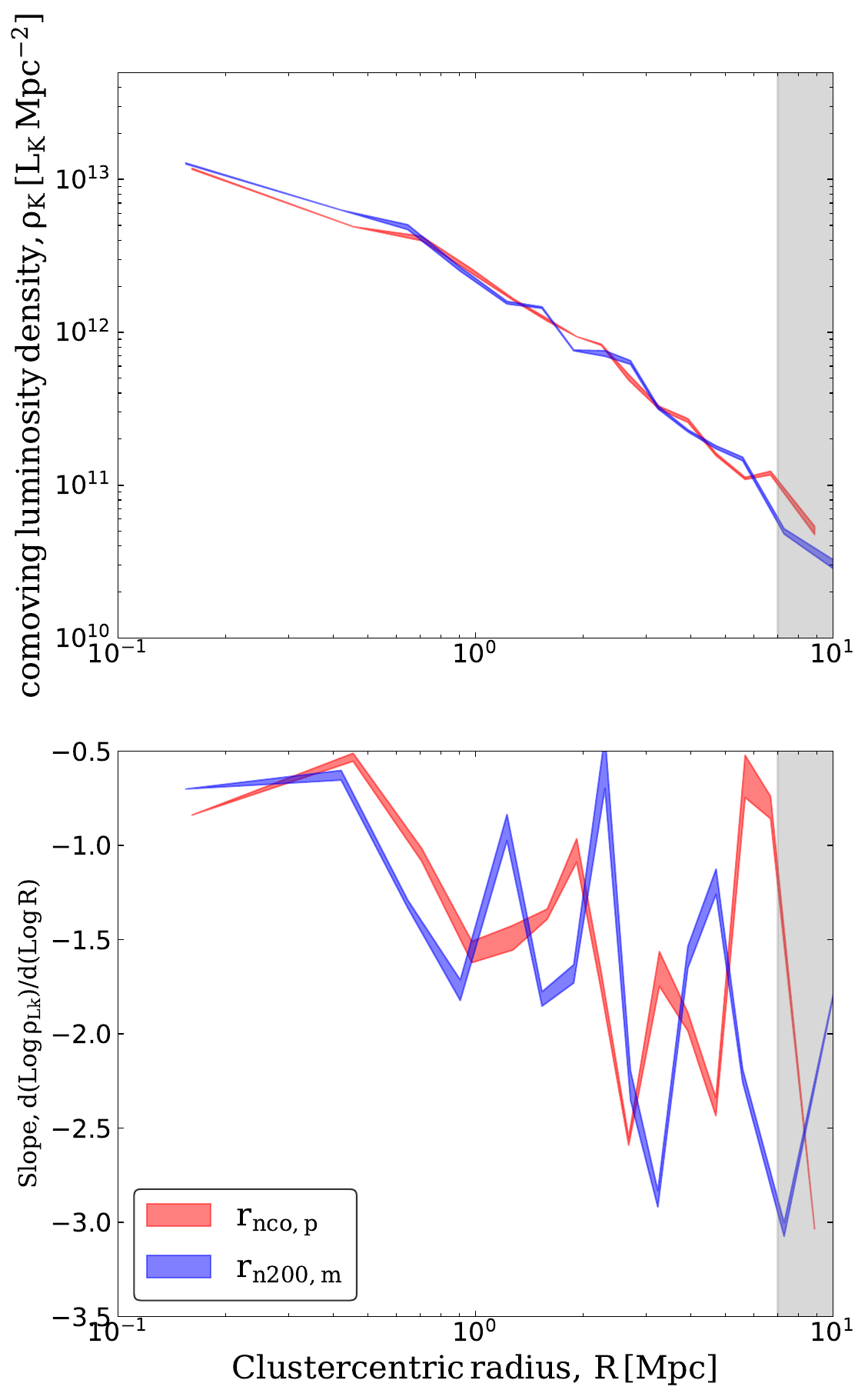} & \includegraphics[width=0.25\linewidth, keepaspectratio]{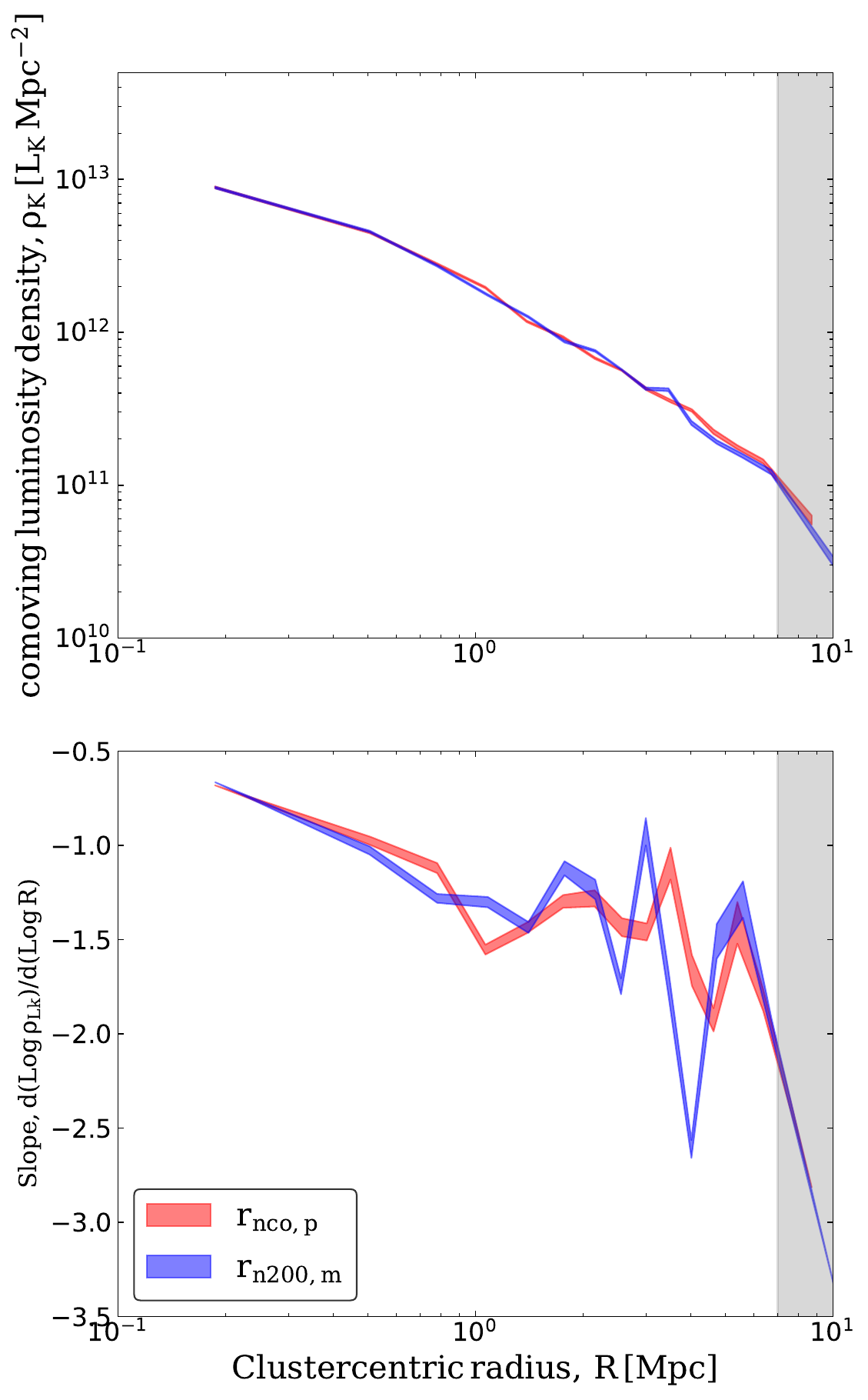} & \includegraphics[width=0.25\linewidth, keepaspectratio]{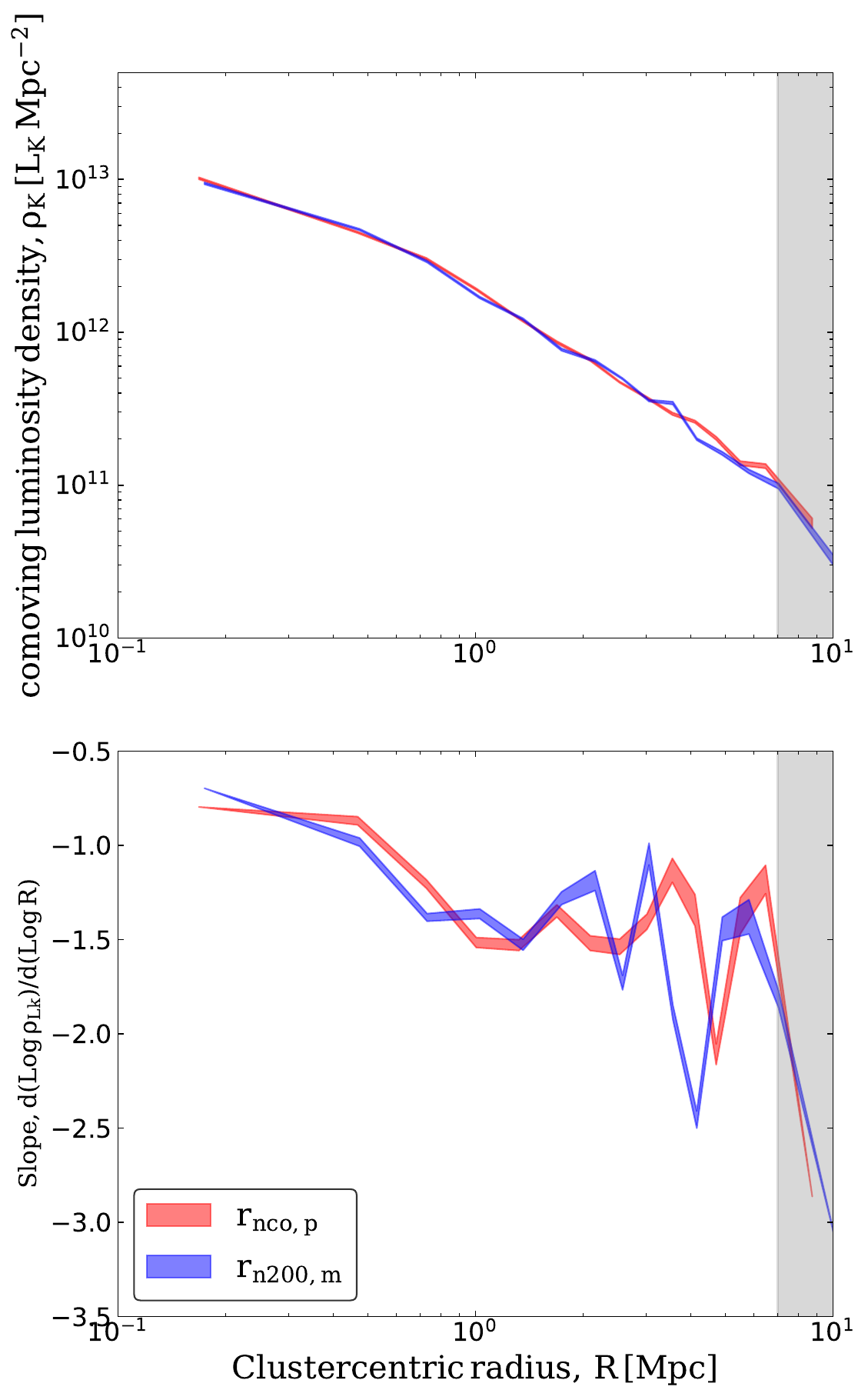} & \includegraphics[width=0.25\linewidth, keepaspectratio]{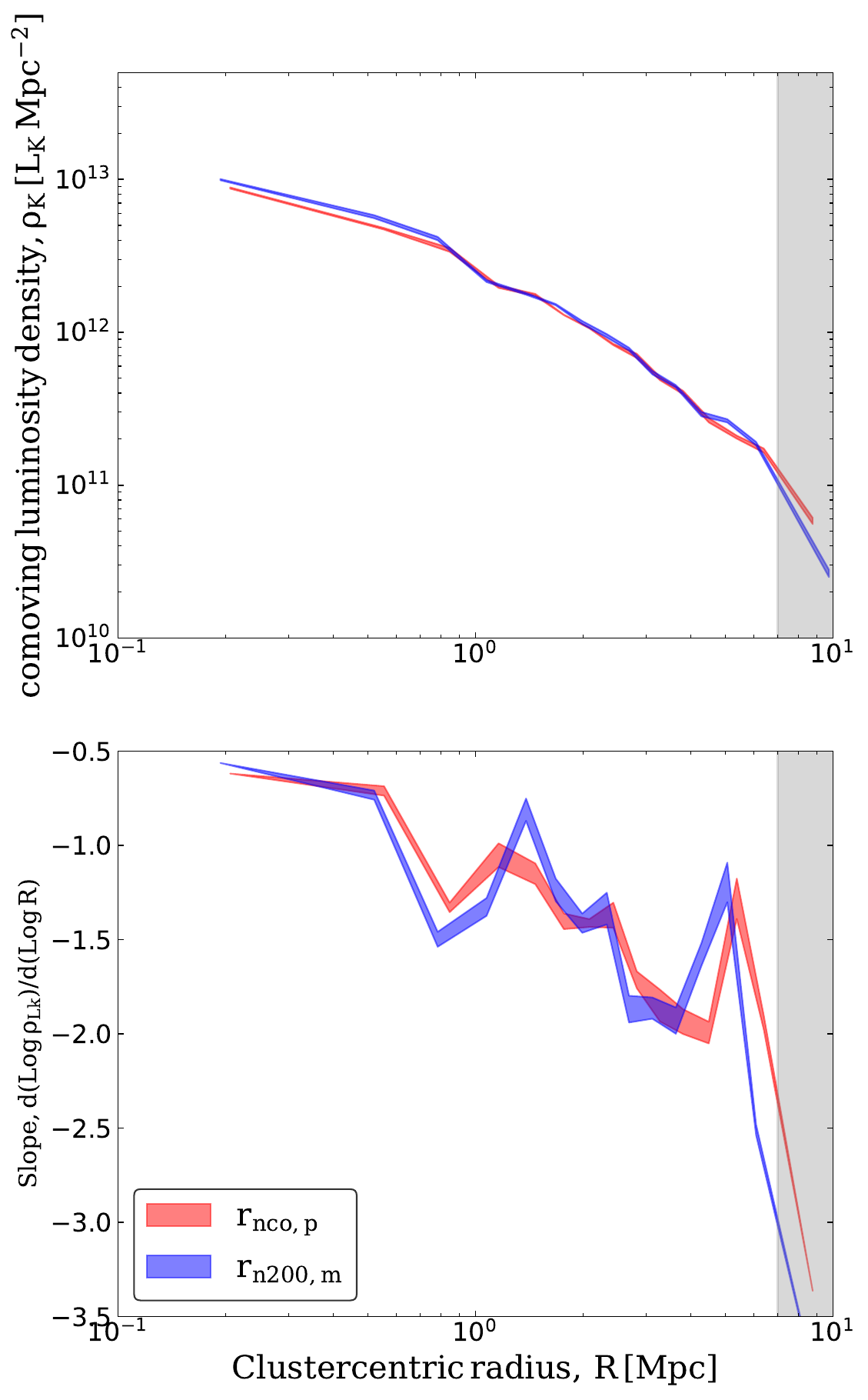}
\end{tabular}

\caption{Continued from Figure~\ref{plot:nrholk_profiles_subsamples}.}\label{plot:nrholk_profiles_subsamples2}
\end{figure*}

\end{document}